\documentclass[twocolumn]{aastex63}

\usepackage{subfigure}
\usepackage{url}
\usepackage{hyperref}
\usepackage{epsfig}
\usepackage{graphicx}
\usepackage{sidecap}
\usepackage{hanging}
\usepackage{color}
\usepackage{multirow}
\usepackage{enumerate}
\usepackage{natbib}
\usepackage{amssymb}
\usepackage{amsmath}
\usepackage[bottom]{footmisc}

\newcommand{\tess}{{TESS}}

\providecommand{\e}[1]{\ensuremath{\, \times \, 10^{#1}}}

\newcommand{\tic}{{TIC~199376584}}
\newcommand{\hd}{{HD~332231}}
\newcommand{\hdb}{{HD~332231~b}}
\newcommand{\toi}{TOI~1456}

\received{2020 January 16}
\revised{2020 March 20}
\accepted{2020 March 27}

\submitjournal{The Astronomical Journal}

\shorttitle{TESS Discovery of a Planet Orbiting HD~332231}
\shortauthors{Dalba et al.}

\begin{document}

\title{The TESS--Keck Survey. I.\\A Warm Sub-Saturn-mass Planet and a Caution about Stray Light in TESS Cameras\footnote{Some of the data presented herein were obtained at the W. M. Keck Observatory, which is operated as a scientific partnership among the California Institute of Technology, the University of California and the National Aeronautics and Space Administration. The Observatory was made possible by the generous financial support of the W. M. Keck Foundation.}}

\correspondingauthor{Paul A. Dalba}
\email{pdalba@ucr.edu}



\author[0000-0002-4297-5506]{Paul A.\ Dalba} 
\altaffiliation{NSF Astronomy \& Astrophysics Postdoctoral Fellow}
\affiliation{Department of Earth \& Planetary Sciences, University of California Riverside, 900 University Ave, Riverside, CA 92521, USA}

\author[0000-0002-5463-9980]{Arvind F.\ Gupta} 
\affiliation{Department of Astronomy \& Astrophysics, 525 Davey Laboratory, The Pennsylvania State University, University Park, PA, 16802, USA}
\affiliation{Center for Exoplanets \& Habitable Worlds, 525 Davey Laboratory, The Pennsylvania State University, University Park, PA, 16802, USA}

\author[0000-0001-8812-0565]{Joseph E.\ Rodriguez} 
\affiliation{Center for Astrophysics \textbar \ Harvard \& Smithsonian, 60 Garden St, Cambridge, MA 02138, USA}

\author[0000-0003-2313-467X]{Diana Dragomir} 
\affiliation{Department of Physics \& Astronomy, University of New Mexico, 1919 Lomas Blvd NE, Albuquerque, NM 87131, USA}

\author[0000-0003-0918-7484]{Chelsea X.\ Huang} 
\altaffiliation{Juan Carlos Torres Fellow}
\affiliation{Department of Physics and Kavli Institute for Astrophysics and Space Research, Massachusetts Institute of Technology, Cambridge, MA 02139, USA}

\author[0000-0002-7084-0529]{Stephen R.\ Kane} 
\affiliation{Department of Earth \& Planetary Sciences, University of California Riverside, 900 University Ave, Riverside, CA 92521, USA}

\author[0000-0002-8964-8377]{Samuel N.\ Quinn} 
\affiliation{Center for Astrophysics \textbar \ Harvard \& Smithsonian, 60 Garden St, Cambridge, MA 02138, USA}

\author[0000-0001-6637-5401]{Allyson Bieryla} 
\affiliation{Center for Astrophysics \textbar \ Harvard \& Smithsonian, 60 Garden St, Cambridge, MA 02138, USA}

\author[0000-0002-9789-5474]{Gilbert A.\ Esquerdo} 
\affiliation{Center for Astrophysics \textbar \ Harvard \& Smithsonian, 60 Garden St, Cambridge, MA 02138, USA}

\author[0000-0003-3504-5316]{Benjamin J.\ Fulton} 
\affiliation{NASA Exoplanet Science Institute, Caltech/IPAC, Pasadena, CA 91125, USA}

\author[0000-0003-3623-7280]{Nicholas Scarsdale} 
\affiliation{Department of Astronomy \& Astrophysics, University of California, Santa Cruz, CA 95064, USA}


\author[0000-0002-7030-9519]{Natalie M.\ Batalha} 
\affiliation{Department of Astronomy \& Astrophysics, University of California, Santa Cruz, CA 95064, USA}

\author[0000-0001-7708-2364]{Corey Beard}
\affiliation{Department of Physics \& Astronomy, The University of California, Irvine, 4129 Frederick Reines Hall, Irvine, CA 92697, USA}

\author[0000-0003-0012-9093]{Aida Behmard}
\altaffiliation{NSF Graduate Research Fellow}
\affiliation{Division of Geological \& Planetary Science, California Institute of Technology, 1200 E California Blvd, Pasadena, CA 91125, USA}

\author[0000-0003-1125-2564]{Ashley Chontos}
\altaffiliation{NSF Graduate Research Fellow}
\affiliation{Institute for Astronomy, University of Hawai`i, 2680 Woodlawn Drive, Honolulu, HI 96822, USA}

\author{Ian J.\ M.\ Crossfield}
\affiliation{Department of Physics \& Astronomy, University of
  Kansas, Lawrence, KS, USA}

\author[0000-0001-8189-0233]{Courtney D.\ Dressing} 
\affiliation{Department of Astronomy, University of California Berkeley, Berkeley, CA 94720-3411, USA}

\author[0000-0002-8965-3969]{Steven Giacalone}
\affiliation{Department of Astronomy, University of California Berkeley, Berkeley, CA 94720-3411, USA}

\author[0000-0002-0139-4756]{Michelle L.\ Hill}
\affiliation{Department of Earth \& Planetary Sciences, University of California Riverside, 900 University Ave, Riverside, CA 92521, USA}

\author[0000-0001-8058-7443]{Lea A.\ Hirsch}
\affiliation{Kavli Institute for Particle Astrophysics and Cosmology, Stanford University, Stanford, CA 94305, USA}

\author[0000-0001-8638-0320]{Andrew W.\ Howard} 
\affiliation{Department of Astronomy, California Institute of Technology, 1200 E California Blvd, Pasadena, CA 91125, USA}

\author[0000-0001-8832-4488]{Daniel Huber} 
\affiliation{Institute for Astronomy, University of Hawai`i, 2680 Woodlawn Drive, Honolulu, HI 96822, USA}

\author[0000-0002-0531-1073]{Howard Isaacson} 
\affiliation{Department of Astronomy, University of California Berkeley, Berkeley, CA 94720-3411, USA}
\affiliation{University of Southern Queensland, Toowoomba, QLD 4350, Australia}

\author[0000-0002-6115-4359]{Molly Kosiarek}
\altaffiliation{NSF Graduate Research Fellow}
\affiliation{Department of Astronomy \& Astrophysics, University of California, Santa Cruz, CA 95064, USA} 

\author[0000-0001-8342-7736]{Jack Lubin}
\affiliation{Department of Physics \& Astronomy, The University of California, Irvine, 4129 Frederick Reines Hall, Irvine, CA 92697, USA}

\author[0000-0002-7216-2135]{Andrew W.\ Mayo}
\affiliation{Department of Astronomy, University of California Berkeley, Berkeley, CA 94720-3411, USA}
\affiliation{Centre for Star \& Planet Formation, Natural History Museum of Denmark \& Niels Bohr Institute, University of Copenhagen, \O{}ster Voldgade 5-7, DK-1350 Copenhagen K, Denmark}

\author[0000-0003-4603-556X]{Teo Mocnik}
\affiliation{Gemini Observatory Northern Operations, 670 N. A'ohoku Place, Hilo, HI 96720, USA}

\author[0000-0001-8898-8284]{Joseph M.\ Akana Murphy}
\altaffiliation{NSF Graduate Research Fellow, LSSTC Data Science Fellow}
\affiliation{Department of Astronomy \& Astrophysics, University of California, Santa Cruz, CA 95064, USA}

\author[0000-0003-0967-2893]{Erik A.\ Petigura} 
\affiliation{Department of Physics \& Astronomy, University of California Los Angeles, Los Angeles, CA 90095, USA}

\author[0000-0003-0149-9678]{Paul Robertson} 
\affiliation{Department of Physics \& Astronomy, The University of California, Irvine, 4129 Frederick Reines Hall, Irvine, CA 92697, USA}

\author[0000-0001-8391-5182]{Lee J.\ Rosenthal}
\affiliation{Department of Astronomy, California Institute of Technology, 1200 E California Blvd, Pasadena, CA 91125, USA}

\author[0000-0001-8127-5775]{Arpita Roy} 
\affiliation{Department of Astronomy, California Institute of Technology, 1200 E California Blvd, Pasadena, CA 91125, USA}

\author[0000-0003-3856-3143]{Ryan A.\ Rubenzahl}
\altaffiliation{NSF Graduate Research Fellow}
\affiliation{Department of Astronomy, California Institute of Technology, 1200 E California Blvd, Pasadena, CA 91125, USA}

\author[0000-0002-4290-6826]{Judah Van Zandt}
\affiliation{Department of Physics \& Astronomy, University of California Los Angeles, Los Angeles, CA 90095, USA}

\author[0000-0002-3725-3058]{Lauren M.\ Weiss} 
\affiliation{Institute for Astronomy, University of Hawai`i, 2680 Woodlawn Drive, Honolulu, HI 96822, USA}


\author[0000-0001-7880-594X]{Emil Knudstrup} 
\affiliation{Stellar Astrophysics Centre, Department of Physics and Astronomy, Aarhus University, Ny Munkegade 120, DK-8000 Aarhus C, Denmark}

\author[0000-0002-9194-8520]{Mads F.\ Andersen} 
\affiliation{Stellar Astrophysics Centre, Department of Physics and Astronomy, Aarhus University, Ny Munkegade 120, DK-8000 Aarhus C, Denmark}

\author[0000-0002-8736-1639]{Frank Grundahl} 
\affiliation{Stellar Astrophysics Centre, Department of Physics and Astronomy, Aarhus University, Ny Munkegade 120, DK-8000 Aarhus C, Denmark}


\author[0000-0003-4554-5592]{Xinyu Yao} 
\affiliation{Department of Physics, Lehigh University, 16 Memorial Drive East, Bethlehem, PA 18015, USA}

\author[0000-0002-3827-8417]{Joshua Pepper} 
\affiliation{Department of Physics, Lehigh University, 16 Memorial Drive East, Bethlehem, PA 18015, USA}

\author[0000-0001-6213-8804]{Steven Villanueva Jr.} 
\altaffiliation{Juan Carlos Torres Fellow}
\affiliation{Department of Physics and Kavli Institute for Astrophysics and Space Research, Massachusetts Institute of Technology, Cambridge, MA 02139, USA}


\author[0000-0002-5741-3047]{David R.\ Ciardi} 
\affiliation{NASA Exoplanet Science Institute, Caltech/IPAC, Pasadena, CA 91125, USA}

\author[0000-0001-5383-9393]{Ryan Cloutier} 
\affiliation{Center for Astrophysics \textbar \ Harvard \& Smithsonian, 60 Garden St, Cambridge, MA 02138, USA}

\author[0000-0003-3988-3245]{Thomas Lee Jacobs} 
\altaffiliation{Amateur Astronomer}
\affiliation{12812 SE 69th Place, Bellevue, WA 98006, USA}

\author[0000-0002-2607-138X]{Martti H.\ Kristiansen} 
\affiliation{DTU Space, National Space Institute, Technical University of Denmark, Elektrovej 327, DK-2800 Lyngby, Denmark}
\affiliation{Brorfelde Observatory, Observator Gyldenkernes Vej 7, DK-4340 T\o{}ll\o{}se, Denmark}

\author[0000-0002-8527-2114]{Daryll M.\ LaCourse} 
\altaffiliation{Amateur Astronomer}
\affiliation{7507 52nd Pl NE, Marysville, WA 98270, USA}

\author[0000-0001-9699-1459]{Monika Lendl} 
\affiliation{Observatoire de Gen{\`e}ve, Universit{\'e} de Gen{\`e}ve, 51 Ch. des Maillettes, 1290 Sauverny, Switzerland}
\affiliation{Space Research Institute, Austrian Academy of Sciences, Schmiedlstr. 6, A-8042 Graz, Austria}

\author[0000-0002-4047-4724]{Hugh P.\ Osborn} 
\affiliation{Center for Space and Habitability, University of Bern, Gesellschaftsstrasse 6, CH-3012 Bern, Switzerland}
\affiliation{Department of Physics and Kavli Institute for Astrophysics and Space Research, Massachusetts Institute of Technology, Cambridge, MA 02139, USA}

\author[0000-0003-0987-1593]{Enric Palle}
\affiliation{Instituto de Astrof\'isica de Canarias (IAC), E-38200 La Laguna, Tenerife, Spain}

\author[0000-0002-3481-9052]{Keivan G.\ Stassun} 
\affiliation{Vanderbilt University, Department of Physics \& Astronomy, 6301 Stevenson Center Ln., Nashville, TN 37235, USA}
\affiliation{Fisk University, Department of Physics, 1000 18th Ave. N., Nashville, TN 37208, USA}

\author[0000-0002-5951-8328]{Daniel J.\ Stevens} 
\altaffiliation{Eberly Fellow}
\affiliation{Department of Astronomy \& Astrophysics, 525 Davey Laboratory, The Pennsylvania State University, University Park, PA, 16802, USA}
\affiliation{Center for Exoplanets \& Habitable Worlds, 525 Davey Laboratory, The Pennsylvania State University, University Park, PA, 16802, USA}


\author[0000-0003-2058-6662]{George R.\ Ricker} 
\affiliation{Department of Physics and Kavli Institute for Astrophysics and Space Research, Massachusetts Institute of Technology, Cambridge, MA 02139, USA}

\author[0000-0001-6763-6562]{Roland Vanderspek} 
\affiliation{Department of Physics and Kavli Institute for Astrophysics and Space Research, Massachusetts Institute of Technology, Cambridge, MA 02139, USA}

\author[0000-0001-9911-7388]{David W.\ Latham} 
\affiliation{Center for Astrophysics \textbar \ Harvard \& Smithsonian, 60 Garden St, Cambridge, MA 02138, USA}

\author[0000-0002-6892-6948]{S. Seager}
\affiliation{Department of Physics and Kavli Institute for Astrophysics and Space Research, Massachusetts Institute of Technology, Cambridge, MA 02139, USA}
\affiliation{Department of Earth, Atmospheric and Planetary Sciences, Massachusetts Institute of Technology, Cambridge, MA 02139, USA}
\affiliation{Department of Aeronautics and Astronautics, MIT, 77 Massachusetts Avenue, Cambridge, MA 02139, USA}

\author[0000-0002-4265-047X]{Joshua N.\ Winn} 
\affiliation{Department of Astrophysical Sciences, Princeton University, 4 Ivy Lane, Princeton, NJ 08544, USA}

\author[0000-0002-4715-9460]{Jon M.\ Jenkins} 
\affiliation{NASA Ames Research Center, Moffett Field, CA 94035, USA}


\author[0000-0003-1963-9616]{Douglas A.\ Caldwell} 
\affiliation{NASA Ames Research Center, Moffett Field, CA 94035, USA}
\affiliation{SETI Institute, Mountain View, CA 94043, USA}

\author[0000-0002-6939-9211]{Tansu Daylan} 
\altaffiliation{Kavli Fellow}
\affiliation{Department of Physics and Kavli Institute for Astrophysics and Space Research, Massachusetts Institute of Technology, Cambridge, MA 02139, USA}

\author[0000-0003-0241-2757]{William Fong}
\affiliation{Department of Physics and Kavli Institute for Astrophysics and Space Research, Massachusetts Institute of Technology, Cambridge, MA 02139, USA}

\author[0000-0003-1748-5975]{Robert F.\ Goeke} 
\affiliation{Department of Physics and Kavli Institute for Astrophysics and Space Research, Massachusetts Institute of Technology, Cambridge, MA 02139, USA}

\author[0000-0003-4724-745X]{Mark E.\ Rose} 
\affiliation{NASA Ames Research Center, Moffett Field, CA 94035, USA}

\author[0000-0002-4829-7101]{Pamela Rowden} 
\affiliation{School of Physical Sciences, The Open University, Milton Keynes MK7 6AA, UK}

\author[0000-0001-5347-7062]{Joshua E.\ Schlieder} 
\affiliation{Exoplanets and Stellar Astrophysics Laboratory, Mail Code 667, NASA Goddard Space Flight Center, Greenbelt, MD 20771, USA}

\author[0000-0002-6148-7903]{Jeffrey C.\ Smith} 
\affiliation{NASA Ames Research Center, Moffett Field, CA 94035, USA}
\affiliation{SETI Institute, Mountain View, CA 94043, USA}

\author[0000-0001-7246-5438]{Andrew Vanderburg} 
\altaffiliation{NASA Sagan Fellow}
\affiliation{Department of Astronomy, The University of Texas at Austin, Austin, TX 78712, USA}


\begin{abstract}

We report the detection of a Saturn-size exoplanet orbiting HD~332231 (TOI~1456) in light curves from the Transiting Exoplanet Survey Satellite (TESS). HD~332231---an F8 dwarf star with a $V$-band magnitude of 8.56---was observed by TESS in Sectors 14 and 15. We detect a single-transit event in the Sector 15 presearch data conditioning (PDC) light curve. We obtain spectroscopic follow-up observations of HD~332231 with the Automated Planet Finder, Keck I, and SONG telescopes. The orbital period we infer from the radial velocity (RV) observations leads to the discovery of another transit in Sector 14 that was masked by PDC due to scattered light contamination. A joint analysis of the transit and RV data confirms the planetary nature of HD~332231~b, a Saturn-size ($0.867^{+0.027}_{-0.025} \; R_{\rm J}$), sub-Saturn-mass ($0.244\pm0.021 \; M_{\rm J}$) exoplanet on a 18.71 day circular orbit. The low surface gravity of HD~332231~b and the relatively low stellar flux it receives make it a compelling target for transmission spectroscopy. Also, the stellar obliquity is likely measurable via the Rossiter--McLaughlin effect, an exciting prospect given the 0.14 au orbital separation of HD~332231~b. The spectroscopic observations do not provide substantial evidence for any additional planets in the HD~332231 system, but continued RV monitoring is needed to further characterize this system. We also predict that the frequency and duration of masked data in the PDC light curves for TESS Sectors 14--16 could hide transits of some exoplanets with orbital periods between 10.5 and 17.5~days.
\end{abstract}


 
\section{Introduction}\label{sec:intro}

NASA's Transiting Exoplanet Survey Satellite (\tess) has already led to the discovery of nearly three dozen planets announced in papers that have been submitted or published. Thousands more are expected, and we are well on our way toward this plethora of planets, as evidenced by the over one thousand planet candidate \tess\ Objects of Interest (TOIs) already found (Guerrero et al., submitted). Much of the follow-up effort has been focused on confirming and characterizing the small planets \citep[e.g.,][]{Winters2019,Huang2018b,Vanderburg2019,Dragomir2019}. Several giant exoplanets have been confirmed as well \citep[e.g.,][]{Huber2019,Jones2019,Rodriguez2019,Wang2019}. However, due to observational biases, the \tess\ primary mission strategy, and the ease of characterizing planets that transit frequently, most of the \tess\ planets announced to date orbit relatively close to their stars. 

Yet the ``bright star advantage'' of \tess\ can also apply to longer-period exoplanets \citep[e.g.,][]{Eisner2020}. Only a few dozen exoplanets with orbital periods greater than 15~days and precisely measured radii and radial velocity (RV) determined masses are currently known\footnote{According to the NASA Exoplanet Archive \url{https://exoplanetarchive.ipac.caltech.edu/}.}. Of these, only six orbit stars with $V$-band magnitude brighter than 10. This deficiency has impeded robust tests of the theories surrounding planetary interiors and formation. For example, Jupiter and Saturn are often included on exoplanet mass-radius relations where the Jovian regime is largely informed by hot Jupiters \citep[e.g.,][]{Chen2017}. A more informative comparison could be made to cooler exoplanets that have not experienced dramatically different evolutionary histories.

\tess\ has the potential to reveal hundreds of new, weakly irradiated planets transiting bright stars, but many of them will show just one transit during the primary mission \citep[e.g.,]{Sullivan2015,Cooke2018,Villanueva2019,Gill2020,Lendl2020}. In this paper, we present \hdb, a transiting warm sub-Saturn-mass planet initially discovered as a single-transit planet candidate through the TOI identification and vetting process described in detail in a separate paper (Guerrero et al., submitted). Briefly, TOIs are identified in the \tess\ short-cadence ``postage stamp'' light curves with the Transit Planet Search tool \citep{Twicken2016} or in the long-cadence full-frame image light curves using the Box Least Squares algorithm \citep{Kovacs2002}. TOIs are then vetted through the use of several diagnostics such as an odd-even transit depth test, a weak secondary eclipse test, a difference image centroid test, and through examination of the light curves obtained with multiple apertures. After the \tess\ discovery of \hdb, an RV follow-up campaign ensued, uncovering a likely orbital period of $\sim 18$ days. Since \tess\ observed \hd\ in Sectors 14 and 15, this period would mean that at least one other transit should have been caught. Upon closer investigation, another transit (matching the RV period) was indeed discovered in a section of the \tess\ light curve that had been masked out prior to the TOI search due to contamination by scattered light. This sequence of events mirrors that described by \citet{Dragomir2019} and \citet{Eisner2020} regarding spacecraft reaction wheel desaturation events (i.e., momentum dumps).

We present the photometric, imaging, and spectroscopic observations of \hd\ in Section \ref{sec:obs}. We detail the joint analysis of the \tess\ light curves and RV observations from multiple observatories in Section \ref{sec:analysis}. Section \ref{sec:disc} contains a discussion of our results, atmospheric characterization prospects, and the possibility of additional companions in the system. In the same Section, we also carry out an investigation into the number of transits potentially occurring during Sectors 14, 15, and 16 light curve portions that were affected by scattered light. We conclude in Section \ref{sec:conc}.


\section{Observations}\label{sec:obs}

Light curves from \tess\ gave the first indication of a planet candidate orbiting \hd. Following this discovery, we acquired reconnaissance spectra, speckle imaging, and Doppler spectra to further characterize the nature of the system. The following sections describe these observations and the initial analysis we conducted as these data were collected. When a planet candidate is identified through a limited number of transits---only one in this case---the interaction between detection, transit modeling, and RV monitoring is critical for narrowing the parameter space of the system's properties. We will describe how rapid RV follow-up refined the orbital period of the planet candidate, thereby revealing the existence of a previously missed transit. These steps drastically reduced the amount of time and resources required to characterize the entire system.


\subsection{TESS Photometry}\label{sec:data_tess}

\tess\ observed \hd\ (also known as \tic\ and \toi) in Sectors 14 and 15 with two-minute cadence. Version 4.0 of the Science Processing Operations Center (SPOC) data pipeline \citep{Jenkins2016} processed the observations of \hd\ and produced light curves that we accessed via the \tess\ archive on the Mikulski Archive for Space Telescopes (MAST)\footnote{\url{https://archive.stsci.edu/tess/}}. A single transit is present in the Sector 15 light curve produced by the SPOC presearch data conditioning (PDC) module, which removes instrumental systematics and corrects the photometric aperture for the stellar scene \citep[e.g.,][]{Stumpe2012,Smith2012,Stumpe2014}. No transits are present in the Sector 14 SPOC PDC light curve. According to the Sectors 14--15 Data Validation report for \hd \citep{Twicken2018,Li2019}, the SPOC automated transit search \citep{Jenkins2002,Jenkins2010b} detected the Sector 15 transit. However, lacking a clear additional transit, the pipeline paired this event with a subtle patch of noisy data in Sector 14. As a result, this target was not classified as a SPOC TOI due to its odd-even transit depth comparison statistic. The MIT Quick Look Pipeline \citep{Huang2018b} also detected the transit in Sector 15. With additional manual vetting, this target was assigned the TOI number 1456.

The SPOC Sector 14 PDC light curve contains two data gaps in addition to the usual gap marking \tess\ perigee. Partially motivated by ongoing Doppler spectroscopy observations (Section \ref{sec:data_rv}), we explored the source of these data gaps. We used the \texttt{Lightkurve} package \citep{Lightkurve2018}\footnote{\url{https://docs.lightkurve.org/}} to decode the light curve ``bitmask'' and determine that scattered light caused the two additional data gaps. We then acquired the target pixel data for \hd\ in Sector 14 from MAST to generate light curves that included the data contaminated by scattered light. In what follows, we describe two analyses of the Sector 14 target pixel data. The first analysis produced a ``first-order'' light curve that still contained certain systematic features but was sufficiently precise to search for planetary transits. This procedure serves as a useful recipe for stars with similar background contamination issues. The second analysis produced a cleaner and fully detrended light curve that we later used in our modeling. 

\begin{figure*}
    \centering
    \includegraphics[width=0.8\textwidth]{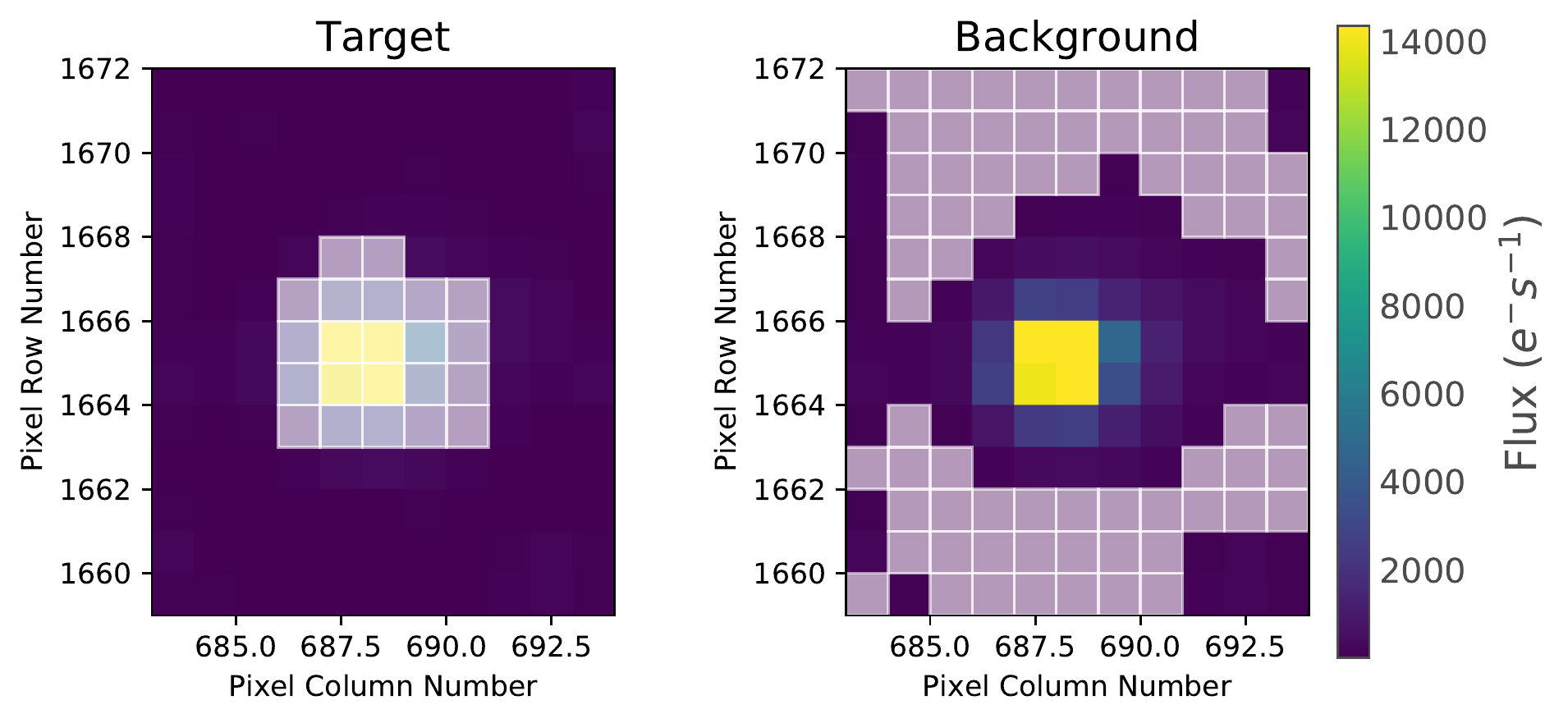}
    \caption{Apertures applied to the Sector 14 target pixel data to generate a ``first-order'' light curve. White boxes indicate pixels included in each aperture. Left: photometric aperture for \hd. Right: background aperture for the correction of scattered light. }
    \label{fig:apertures}
\end{figure*}

\begin{figure}
    \centering
    \includegraphics[width=\columnwidth]{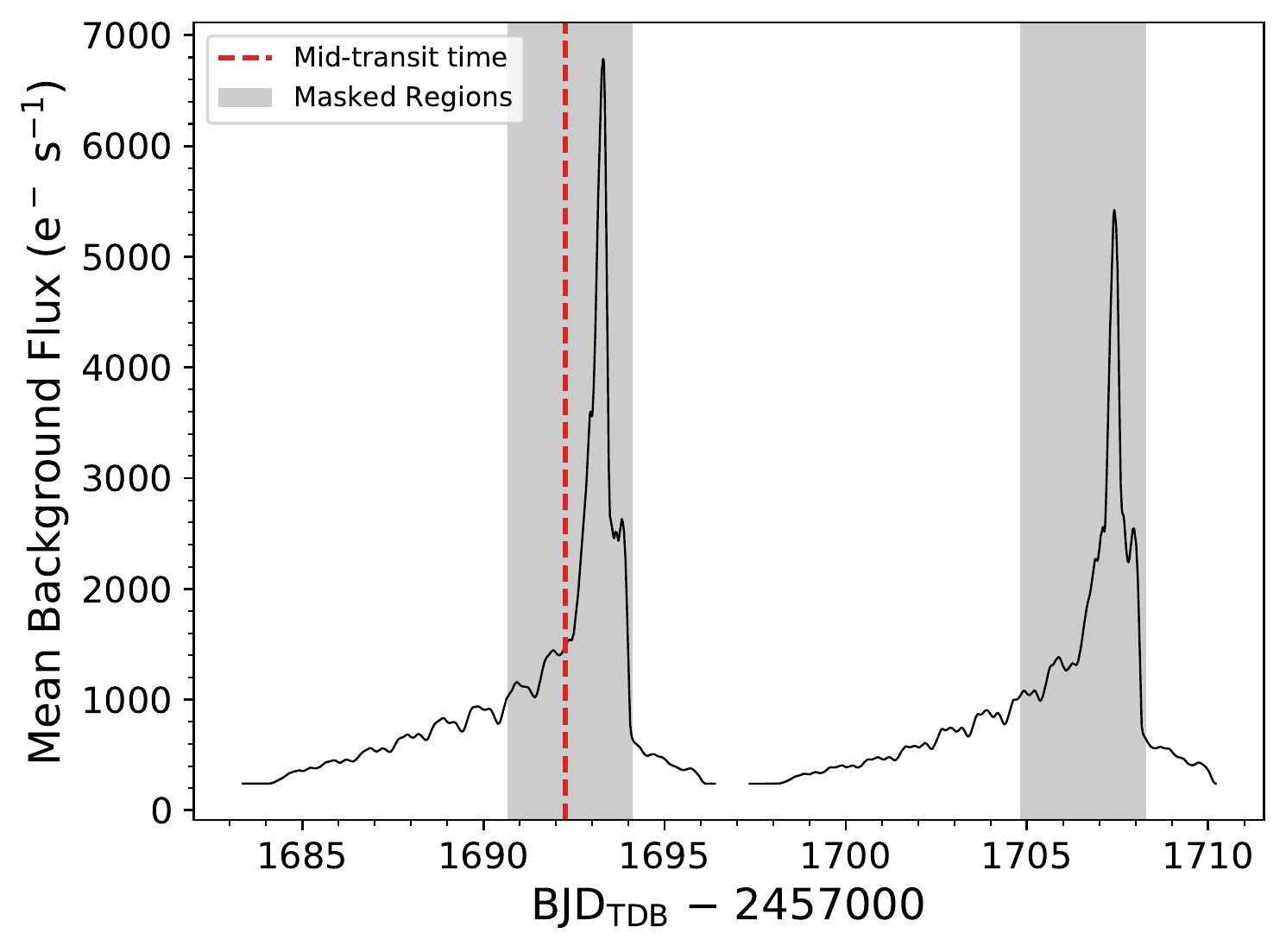}
    \caption{Mean background flux per pixel in the ``first-order'' background aperture for \hd\ in Sector 14. Gray regions highlight data that were masked in the SPOC PDC light curve. The red line marks the mid-time of the transit that was initially hidden due to this masking.}
    \label{fig:S14_scattered}
\end{figure}

To generate the ``first-order'' light curve, we isolated \hd\ between pixel rows 683--694 and pixel columns 1659--1672 (Figure \ref{fig:apertures}). The photometric aperture comprised any pixels in this array with electron fluxes greater than the overall median plus four times the standard deviation. This threshold yielded a roughly rectangular photometric aperture centered on \hd\ consisting of 22 pixels. The background aperture comprised any pixels with electron fluxes below the overall median plus 0.001~times the standard deviation. This threshold is meant to capture all pixels with flux values on the order of the median. Figure \ref{fig:S14_scattered} shows the average background flux as function of time in Sector 14 using this aperture. Indeed, scattered light substantially impacted the observations of \hd.

\begin{figure*}
    \centering
    \includegraphics[width=\textwidth]{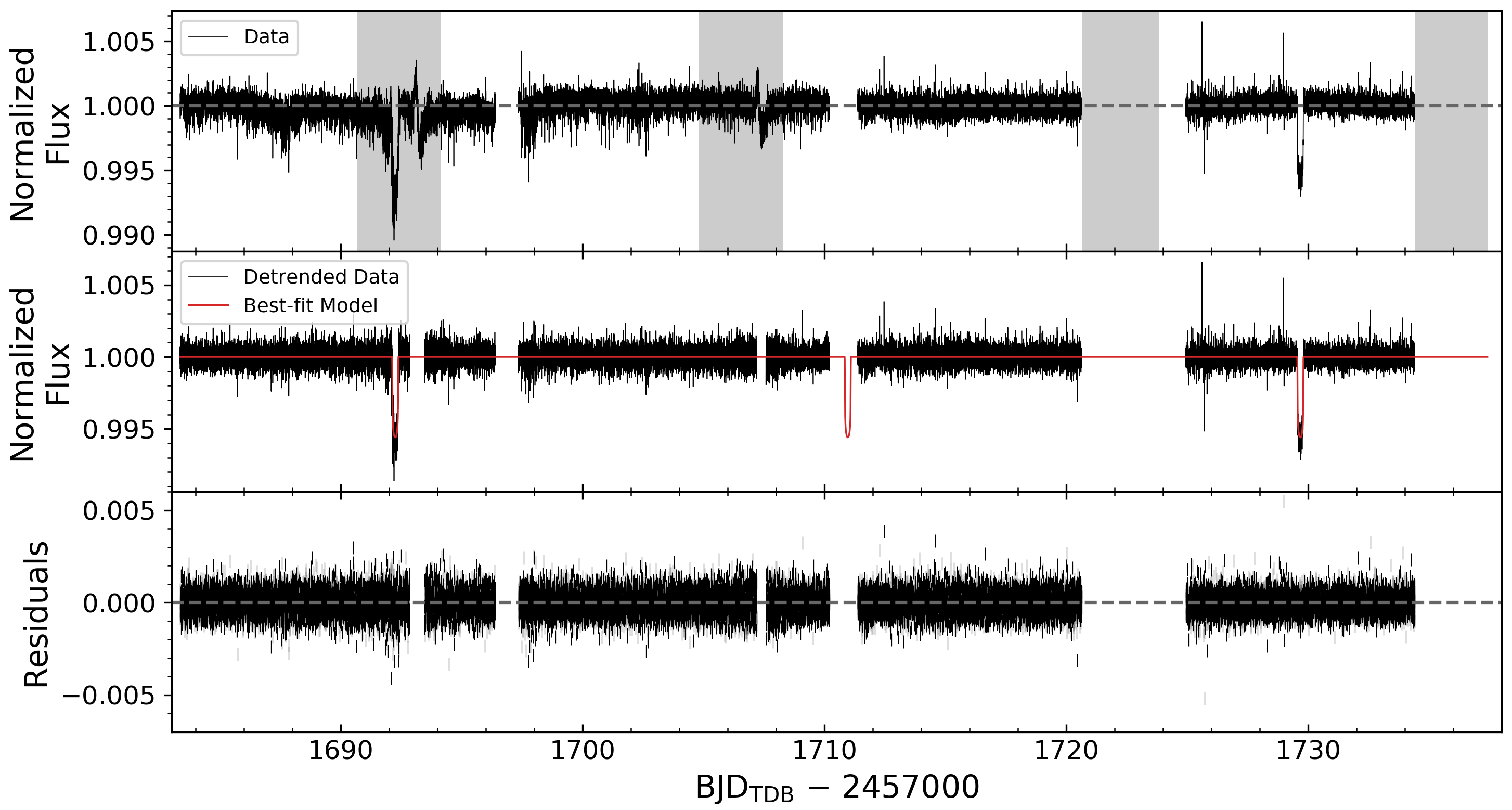}
    \caption{\tess\ photometry of \hd\ from Sectors 14 and 15. Top: normalized light curves prior to detrending. This includes the ``first-order'' light curve from Sector 14 that was extracted from target pixel data using a custom aperture and scattered light subtraction routine (see text). The Sector 15 light curve is the SPOC PDC data product. The gray regions show the timing of the PDC data gaps from stray light masking. Middle: fully detrended light curves and the best-fit model from the global fit (see Section \ref{sec:analysis}). An additional transit of \hdb\ occurred (but was not observed) during the data gap near BJD$_{\rm TDB}\approx$ 2458711, which separates Sectors 14 and 15. Bottom: residuals between the data and best-fit model from the global fit.}
    \label{fig:transit}
\end{figure*}

We then subtracted the background flux time series from the target flux time series (correcting for the difference in aperture sizes) and generated a normalized, ``first-order'' light curve of \hd\ in Sector 14 (Figure \ref{fig:transit}, top panel). In determining the flux in this custom aperture, we did not correct for star field crowding. According to the \tess\ Input Catalog version 8 \citep{Stassun2019}, the brightest source within $100\arcsec$ of \hd\ is 5.63 mag fainter in the \tess\ bandpass. Therefore, it contributes less than 0.6\% of the flux in the photometric aperture. 

A transit event is clearly present in the Sector 14 ``first-order'' light curve generated from the target pixel data. By removing the scattered light using a background aperture, we have uncovered this transit and ensured that it is not diluted by scattered light. This is critical since scattered light could otherwise influence the planetary radius inferred from the transit. 

Having discovered an additional transit event, we re-extracted the Sector 14 light curve, experimenting with photometric apertures of different shapes and sizes, and settled on a light curve from a slightly smaller aperture than that shown in Figure \ref{fig:apertures}. We used a similar strategy to measure and remove scattered background light. We removed instrumental systematics by decorrelating the light curve with the standard deviation of the spacecraft quaternion time series within each exposure \citep{Vanderburg2019} and the median flux value from our background aperture (excluding the in-transit points from the fit). We removed any stellar variability or long-term instrumental trends by dividing a best-fit basis spline from the light curve. The middle panel of Figure \ref{fig:transit} displays the final, flattened Sector 14 light curve of \hd. The standard deviation of the out-of-transit normalized flux is 585~parts per million (ppm), which we treat as the uncertainty on each data point.

\begin{figure}
    \centering
    \includegraphics[width=\columnwidth]{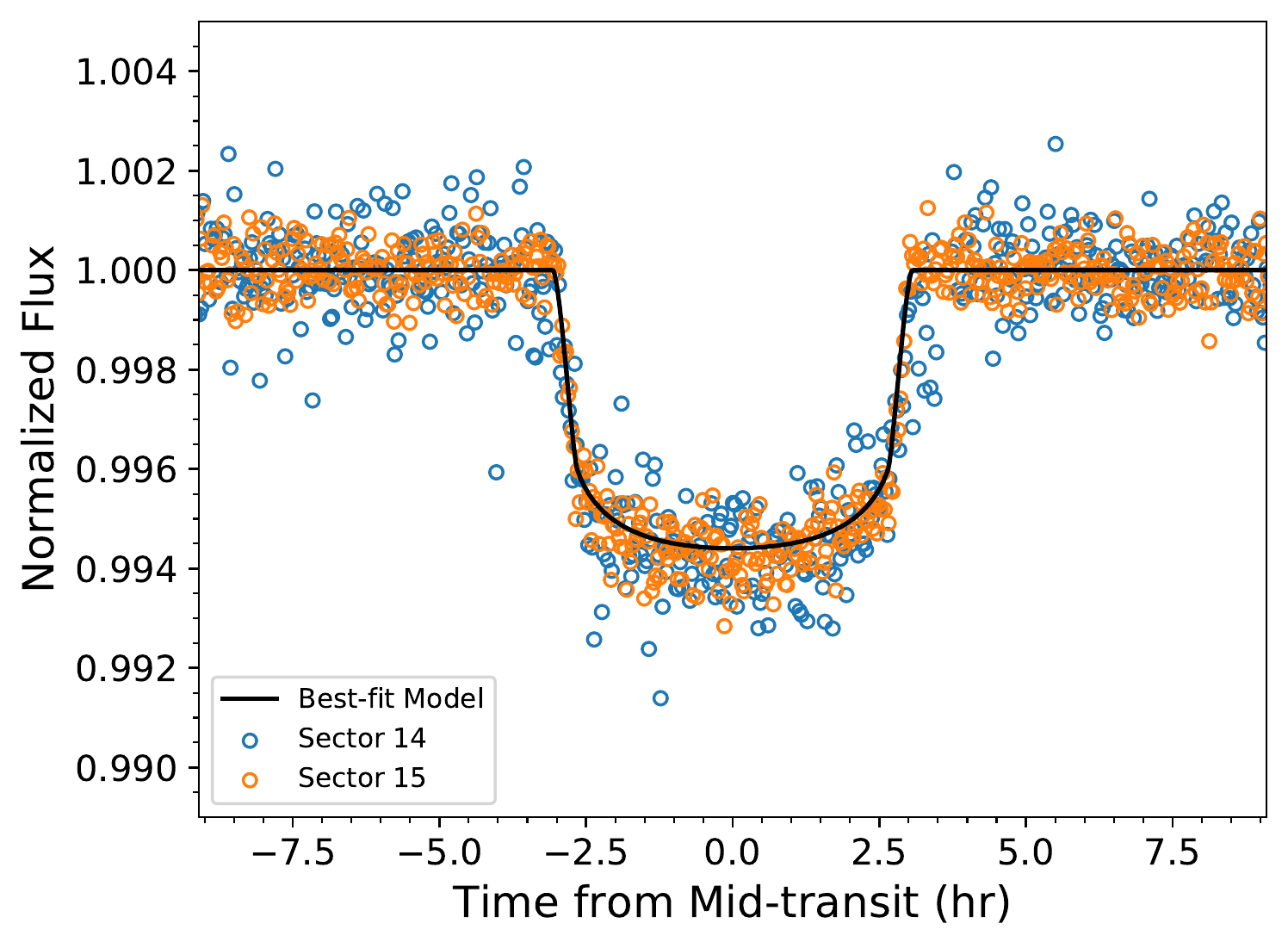}
    \caption{Transits of \hdb\ from Sectors 14 and 15 folded on the best-fit ephemeris from the global model fit (Table \ref{tab:exofastv2}). The best-fit transit model is also shown.}
    \label{fig:transit_fold}
\end{figure}

The PDC light curve of \hd\ from Sector 15 also contains masked data due to scattered light contamination. These data gaps occur near the end of each orbit and do not hide any exoplanet transits. As a result, we proceeded with the SPOC PDC light curve for Sector 15, ignoring the masked data (Figure \ref{fig:transit}, top panel). Any background flux, including scattered light, is removed by the photometric analysis (PA) module prior to PDC \citep{Twicken2010}. Therefore, the Sector 15 transit is not diluted. We flattened the light curve using a Savitzky--Golay filter \citep{Savitzky1964} with a window length of 1001 cadences to remove low-frequency noise. The middle panel of Figure \ref{fig:transit} displays the final, flattened Sector 15 light curve of \hd. The standard deviation of the out-of-transit normalized flux is 547~ppm, which we treat as the uncertainty on each data point. In Figure \ref{fig:transit_fold}, the transit events from Sectors 14 and 15 are plotted together to demonstrate the similarity in depth, duration, and shape. 

\subsection{Reconnaissance Spectroscopy}\label{sec:data_recon}

The \tess\ observations of \hd\ motivated follow-up spectroscopy to characterize the system and explore false-positive scenarios for the transits. We acquired a single reconnaissance spectrum with the Tillinghast Reflector Echelle Spectrograph (TRES) on the 1.5~m telescope at the Fred L. Whipple Observatory (FLWO). The spectrum has a resolution of $R=44000$ in the wavelength range 3850--9096~\AA\ and a signal-to-noise ratio (S/N) of roughly 50. By visual inspection, the TRES spectrum does not contain two sets of spectral lines that would be indicative of a spectroscopic binary system. We used the Stellar Parameter Classification Tool \citep[SPC;][]{Buchhave2012,Buchhave2014} to extract various properties of \hd\ from the TRES spectrum including effective temperature ($T_{\rm eff}$), surface gravity ($\log{g}$), metallicity ($\rm [M/H]$), and projected rotational velocity ($v\sin{i}$). The spectroscopic parameters inferred with SPC are listed in Table \ref{tab:spec}. 

\begin{deluxetable}{lccc}
\tablecaption{Spectroscopic Parameters for \hd\ from TRES, Automated Planet Finder, and Keck Spectra}
\tablehead{
  \colhead{Property} & 
  \colhead{Keck} &
  \colhead{APF} &
  \colhead{TRES}}
\startdata
$T_{\rm eff}$ (K)         & $6100\pm100$  & $6070\pm100$  & $6130\pm50$  \\
$\log g$ (dex)            & $4.3\pm0.1$ & $4.2\pm0.1$   & $4.3\pm0.1$ \\
$\rm [Fe/H]$ (dex)        & $0.04\pm0.06$ & $0.05\pm0.06$ & \nodata \\
$\rm [M/H]$ (dex)         & \nodata       & \nodata       & $-0.01\pm0.08$ \\
$v \sin{i}$ (km s$^{-1}$) & $5.3\pm1.0$   & $5.4\pm1.0$       & $7.0\pm0.5$ \\
$M_{\star}$ ($M_{\sun}$)  & $1.13\pm0.05$ & $1.13\pm0.07$ & \nodata \\
$R_{\star}$ ($R_{\sun}$)  & $1.29\pm0.03$ & $1.4\pm0.2$   & \nodata \\
\enddata
\tablenotetext{}{Note: The final spectroscopic properties that should be adopted for \hd\ derive from the global fit to the transit and RV data (Section \ref{sec:analysis}) and are listed in Table \ref{tab:exofast_stellar}.}
\label{tab:spec}
\end{deluxetable}


\subsection{Speckle Imaging}\label{sec:data_image}

\begin{figure}\label{fig:speckle}
    \centering
    \includegraphics[width=\columnwidth]{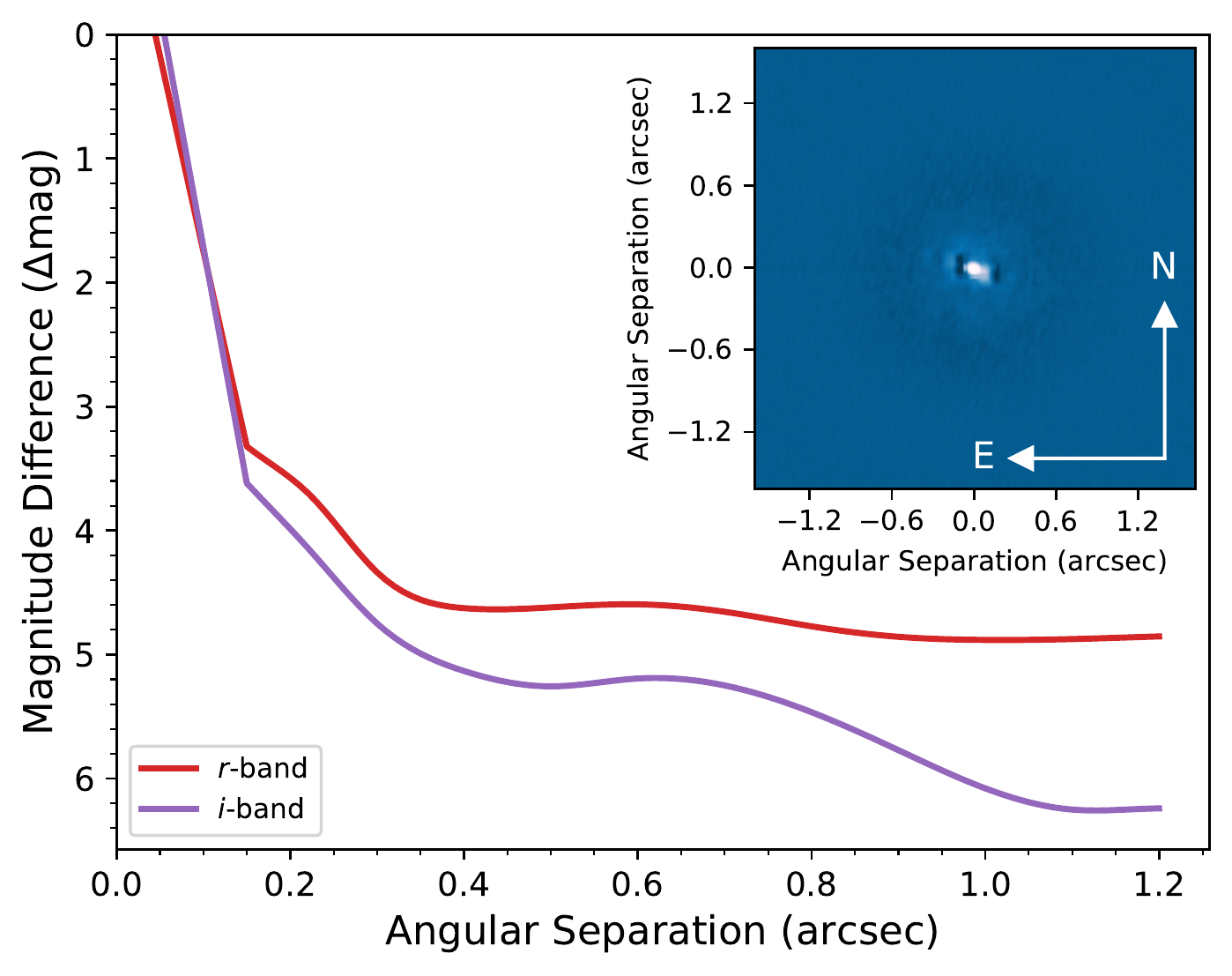}
    \caption{\hd\ contrast curve and reconstructed $i$-band image from the NESSI speckle observations.}
    \label{fig:image}
\end{figure}

\hd\ was observed on the night of 2019 November 9 using the NN-explore Exoplanet Stellar Speckle Imager \citep[NESSI;][]{Scott2018} on the 3.5~m Wisconsin-Indiana-Yale-NOAO (WIYN) Telescope at Kitt Peak National Observatory. Sequences of diffraction-limited, 40 ms speckle exposures were taken simultaneously in the NESSI $r$-band and the NESSI $i$-band on the blue and red cameras, respectively. The image was reconstructed following the procedures outlined in \citet{Howell2011}. In Figure \ref{fig:speckle}, we show that we achieve magnitude contrasts of $\Delta i > 5$ and $\Delta r > 4$ at a separation of less than $0\farcs4$. The speckle image rules out the possibility that the transit signal observed by \tess\ was due to a faint stellar companion or a blended eclipsing binary.

\subsection{Doppler Spectroscopy}\label{sec:data_rv}

We began monitoring the RV variation of \hd\ in 2019 October to measure the companion's mass and further characterize the system. The majority of the RV observations were acquired with the 2.4~m Automated Planet Finder Telescope \citep[APF;][]{Radovan2014,Vogt2014} at Lick Observatory under the TESS--Keck Survey (TKS) and another program dedicated to single-transit follow-up (2019B-A006, PI: Dalba). TKS is a collaboration between the University of California, the California Institute of Technology, the University of Hawaii, and NASA organized through the California Planet Search with the goal of acquiring substantial RV follow-up observations for \tess. We supplemented the APF data set with additional observations from the 10~m Keck I telescope at the W. M. Keck Observatory. Here, we describe the collection and analysis of each set of RV measurements.


\subsubsection{APF-Levy}

The APF telescope is equipped with the Levy Spectrograph, which is a high-resolution slit-fed optical echelle spectrometer \citep{Radovan2010}. APF-Levy observed \hd\ using the W decker, which has a width of 1$\arcsec$ in the spectral dimension and 3$\arcsec$ in the spatial dimension. This slit provides a resolving power of $R\approx$114000 for a continuous spectrum between 3740 and 9700~\AA. A cell containing warm iodine gas placed in the light path enables the precise wavelength calibration of the observed spectra and instrument profile tracking. Additionally, we acquired a high-S/N spectrum of \hd\ without the iodine cell which was used as a template for measuring the Doppler shifts of spectral absorption lines. During each exposure, an exposure meter monitored guider images to determine the photon-weighted mid-exposure time and when the desired precision had been reached \citep{Kibrick2006}. We extracted a relative RV from each spectrum following the forward modeling process described by \citet{Fulton2015a} and originally designed by \citet{Butler1996}. Furthermore, we calculated the stellar chromospheric activity $S_{\rm HK}$-index for each exposure following \citet{Isaacson2010}. The Pearson correlation coefficient between the $S_{\rm HK}$-indices and the RV measurements is 0.032, which suggests little to no correlation between them. 

\begin{figure}
    \centering
    \includegraphics[width=\columnwidth]{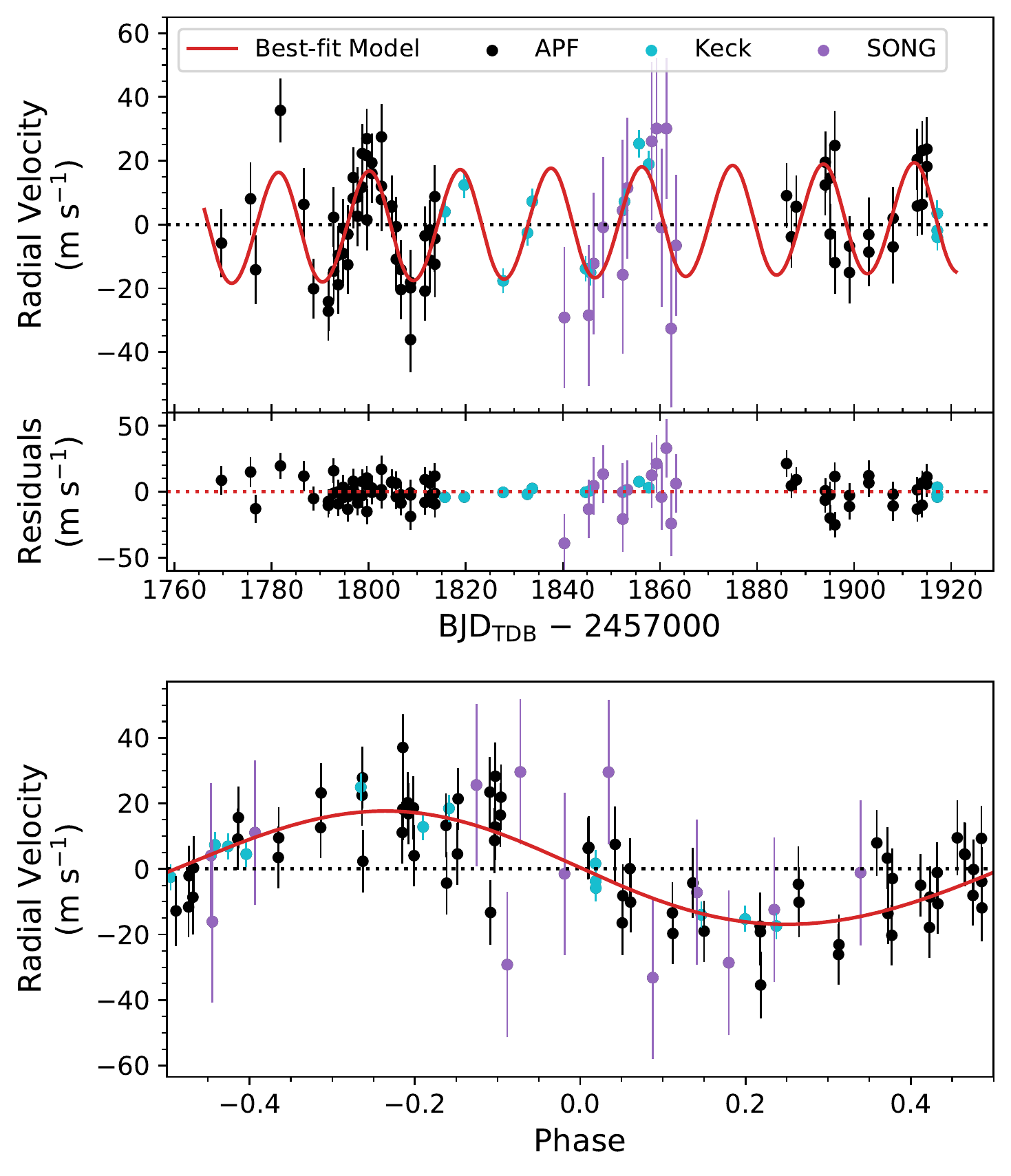}
    \caption{RV observations of \hd\ and best-fit model described in Section \ref{sec:analysis}. Top: the time-series RV observations from APF-Levy, Keck-HIRES, and SONG clearly identify a Keplerian signal due to \hdb. The best-fit model includes an acceleration parameter that describes the slope of the RVs, but such a trend is not statistically preferred by the data (see Section \ref{sec:disc_comp} for a discussion). Bottom: RV observations of \hd\ folded on the best-fit orbital ephemeris.}
    \label{fig:rvs}
\end{figure}

For most of the APF-Levy observations, we acquired two or three successive exposures to increase signal while keeping each exposure time shorter than 20~minutes. The APF telescope is robotic and operates according to a queue that determines suitable targets based on priority and observing conditions. However, on rare occasions, \hd\ was observed in sub-optimal conditions. We manually excluded four exposures from our analysis that were acquired during civil twilight or with seeing greater than 3\arcsec or that had an S/N from photon statistics below 25. In total, APF-Levy collected 68 science-quality iodine-in spectra of \hd\ (Table \ref{tab:rvs}). The time series and phase-folded RV observations of \hd\ are shown in Figure \ref{fig:rvs}.

\begin{deluxetable}{cccc}
\tabletypesize{\footnotesize}
\tablecaption{RV Measurements of \hd}
\label{tab:rvs}
\tablehead{
  \colhead{BJD$_{\rm TDB} - $} & 
  \colhead{RV} &
  \colhead{$S_{\rm HK}$\tablenotemark{a}} &
  \colhead{Tel.}\\
  \colhead{2457000} &
  \colhead{(m s$^{-1}$)} &
  \colhead{} &
  \colhead{}}
\startdata
1769.65399 & $-4.6\pm6.1$ & $0.101\pm0.002$ & APF \\
1775.63931 & $9.4\pm7.3$ & $0.152\pm0.002$ & APF \\
1776.66888 & $-12.9\pm6.1$ & $0.147\pm0.002$ & APF \\
1781.80269 & $37.1\pm4.9$ & $0.147\pm0.002$ & APF \\
1786.60870 & $7.6\pm7.4$ & $0.121\pm0.002$ & APF \\
\enddata
\tablenotetext{a}{The $S_{\rm HK}$ values measured by APF-Levy and Keck-HIRES have different zero-points.}
\tablecomments{Table \ref{tab:rvs} is published in its entirety in the machine-readable format. A portion is shown here for guidance regarding its form and content.}
\end{deluxetable}

Since the high-S/N template spectrum of \hd\ was not observed through the iodine cell, we extracted stellar spectroscopic parameters from this spectrum following the synthetic spectral matching tool \texttt{SpecMatch} \citep{Fulton2015b,Petigura2017}. Stellar parameters output from this procedure are shown in Table \ref{tab:spec}. For parameters that were also extracted from the TRES spectrum of \hd\ (Section \ref{sec:data_recon}), we find agreement with the APF-Levy values to better than $2\sigma$.

While the Doppler monitoring of \hd\ with the APF was ongoing, we modeled the RV variation and inferred planetary and orbital parameters of \hdb\ using the \texttt{RadVel} package\footnote{\url{https://radvel.readthedocs.io/en/latest/}} \citep{Fulton2018}. \texttt{RadVel} samples the posterior probability densities of the various model parameters using the Markov chain Monte Carlo sampler \texttt{emcee} \citep{ForemanMackey2013}. After the first dozen RV epochs, we began to suspect that the orbital period of \hdb\ was short enough to have placed another transit in Sector 14. This inspired the manual analysis of target pixel level data described in Section \ref{sec:data_tess}.


\subsubsection{Keck-HIRES}

We supplemented the Doppler observations of \hd\ from APF-Levy with RVs obtained at the Keck I telescope. Keck I is equipped with the High Resolution Echelle Spectrometer \citep[HIRES;][]{Vogt1994}, which has a substantial record of characterizing exoplanet systems \citep[e.g.,][]{Howard2010,Fulton2017}. We acquired 13 high-resolution ($R \approx 60000$) spectra of \hd. As for APF-Levy, the starlight passed through a heated iodine cell that enabled precise wavelength calibration. We also acquired a high-S/N spectral template using Keck-HIRES that was used in the forward modeling procedure to determine the relative RVs of \hd\ \citep{Howard2010,Howard2016}. The Keck-HIRES RVs for \hd\ are given in Table \ref{tab:rvs} and are shown in Figure \ref{fig:rvs}. Each RV measurement from Keck-HIRES is also accompanied by an S$_{\rm HK}$-index \citep{Isaacson2010}. Importantly, the S$_{\rm HK}$-values measured in the Keck-HIRES spectra have different zero-points than those measured in the APF-Levy spectra. They should not be directly compared. The Pearson correlation coefficient between the Keck-HIRES $S_{\rm HK}$-indices and RVs is 0.29. We caution, however, that this metric is hard to interpret with so few data points. The lack of correlation between the APF $S_{\rm HK}$-indices and the RVs for \hd\ increases the confidence that the Keplerian signal is not an artifact of stellar chromospheric activity. Just as with the APF high-S/N template, we processed the Keck-HIRES template with \texttt{SpecMatch}. The resulting spectroscopic parameters are listed in Table \ref{tab:spec} for comparison with similar parameters derived from APF and TRES observations. The spectroscopic parameters inferred from the spectra from each of these three sources are consistent to well within $2\sigma$ in all cases.


\subsubsection{SONG Telescope}

Additional RV measurements were collected using the robotic 1~m Hertzsprung Stellar Observations Network Group (SONG) telescope \citep[e.g.,][]{Andersen2014,FredslundAndersen2019} which is equipped with a high-resolution echelle spectrograph and uses the iodine method for precise RV calibration \citep{Grundahl2017}. Thirteen spectra of \hd\ were acquired using a slit of width 1\farcs24 ($R\approx90000$) covering the wavelength range 4400--6900~\AA. Each spectrum was obtained using an exposure time of 2700~s resulting in an S/N of $\sim$50. The RVs were determined using the iSONG pipeline \citep[e.g.,][]{Corsaro2012,Antoci2013}, which requires a high-S/N template spectrum ($R\approx110000$). The spectral-line-spread function of the spectrograph was determined by observing a fast rotating O star (HR~1087) through the iodine cell. Uncertainties range from 10~m~s$^{-1}$ for epochs with a high S/N to 15~m~s$^{-1}$ for those with a lower S/N. The SONG RVs are given in Table \ref{tab:rvs} and displayed in Figure \ref{fig:rvs}.


\section{Global Modeling}\label{sec:analysis}
Using the \texttt{EXOFASTv2} modeling suite \citep{Eastman2013,Eastman2019}, we globally fit the available RVs from APF-Levy and Keck-HIRES with the \tess\ photometry to extract the full system parameters. We first conduct a fit to the archival broadband photometry with a model spectral energy distribution (SED) of \hd\ to get a constraint on $R_{\star}$. For this SED-only fit, we use the MESA Isochrones and Stellar Tracks (MIST) stellar evolution models \citep{Dotter2016,Choi2016, Paxton2011,Paxton2013,Paxton2015} and a Gaussian prior on $\rm [Fe/H]$ (0.04$\pm$0.06 dex) from the \texttt{SpecMatch} analysis of the Keck-HIRES template spectrum, and on the parallax measured by Gaia ($12.4498\pm0.04404$~mas), corrected for the offset reported by \citet{Stassun2018a}. We also place an upper limit on the maximum line-of-extinction from the galactic reddening maps from \citet{Schlegel1998}. We get a radius of 1.276 R$_{\odot}$ from this analysis. We then globally fit the available RVs with the \tess\ photometry, and place Gaussian priors on $T_{\rm eff}$ (6100$\pm$100 K) and $\rm [Fe/H]$ (0.04$\pm$0.06 dex) from spectroscopy. We use the determined radius from the SED fit with a 3.5\% prior as an additional Gaussian prior on the fit. We adopt this prior since interferometric radii measurements can show differences of up to $\sim$3\% \citep{White2018}. Within this fit, the MIST isochrones are used to determine the stellar parameters in combination with the priors listed above (Table \ref{tab:exofast_stellar}). The results of the fit are shown in Figures \ref{fig:transit}, \ref{fig:transit_fold}, and \ref{fig:rvs}. A set of fitted and derived planetary parameters for \hdb\ are listed in Table \ref{tab:exofastv2}.

\providecommand{\bjdtdb}{\ensuremath{\rm {BJD_{TDB}}}}
\providecommand{\feh}{\ensuremath{\left[{\rm Fe}/{\rm H}\right]}}
\providecommand{\teff}{\ensuremath{T_{\rm eff}}}
\providecommand{\ecosw}{\ensuremath{e\cos{\omega_*}}}
\providecommand{\esinw}{\ensuremath{e\sin{\omega_*}}}
\providecommand{\msun}{\ensuremath{\,M_\Sun}}
\providecommand{\rsun}{\ensuremath{\,R_\Sun}}
\providecommand{\lsun}{\ensuremath{\,L_\Sun}}
\providecommand{\mj}{\ensuremath{\,M_{\rm J}}}
\providecommand{\rj}{\ensuremath{\,R_{\rm J}}}
\providecommand{\me}{\ensuremath{\,M_{\rm E}}}
\providecommand{\re}{\ensuremath{\,R_{\rm E}}}
\providecommand{\fave}{\langle F \rangle}
\providecommand{\fluxcgs}{10$^9$ erg s$^{-1}$ cm$^{-2}$}
\begin{table}
\scriptsize
\centering
\caption{Median values and 68\% confidence intervals for stellar parameters in the global model of HD~332231}
\begin{tabular}{llcccc}
  \hline
  \hline
Parameter & Description (Units) & Values & & & \\
\hline
~~~~$M_*$\dotfill &Mass (\msun)\dotfill &$1.127\pm0.077$\\
~~~~$R_*$\dotfill &Radius (\rsun)\dotfill &$1.277^{+0.039}_{-0.036}$\\
~~~~$L_*$\dotfill &Luminosity (\lsun)\dotfill &$2.02^{+0.19}_{-0.18}$\\
~~~~$\rho_*$\dotfill &Density (g~cm$^{-3}$)\dotfill &$0.766^{+0.061}_{-0.072}$\\
~~~~$\log{g}$\dotfill &Surface gravity\dotfill &$4.279^{+0.027}_{-0.034}$\\
~~~~$T_{\rm eff}$\dotfill &Effective temperature (K)\dotfill &$6089^{+97}_{-96}$\\
~~~~$[{\rm Fe/H}]$\dotfill &Metallicity (dex)\dotfill &$0.036^{+0.059}_{-0.058}$\\
~~~~$[{\rm Fe/H}]_{0}^\dagger$\dotfill &Initial metallicity \dotfill &$0.095^{+0.055}_{-0.057}$\\
~~~~$Age$\dotfill &Age (Gyr)\dotfill &$4.3^{+2.5}_{-1.9}$\\
\hline
\end{tabular}
\begin{flushleft} 
  \footnotesize{ 
$^\dagger$The initial metallicity is the metallicity of the star when it was formed.}
 \end{flushleft}
\label{tab:exofast_stellar}
\end{table}

\providecommand{\bjdtdb}{\ensuremath{\rm {BJD_{TDB}}}}
\providecommand{\feh}{\ensuremath{\left[{\rm Fe}/{\rm H}\right]}}
\providecommand{\teff}{\ensuremath{T_{\rm eff}}}
\providecommand{\ecosw}{\ensuremath{e\cos{\omega_*}}}
\providecommand{\esinw}{\ensuremath{e\sin{\omega_*}}}
\providecommand{\msun}{\ensuremath{\,M_\Sun}}
\providecommand{\rsun}{\ensuremath{\,R_\Sun}}
\providecommand{\lsun}{\ensuremath{\,L_\Sun}}
\providecommand{\mj}{\ensuremath{\,M_{\rm J}}}
\providecommand{\rj}{\ensuremath{\,R_{\rm J}}}
\providecommand{\me}{\ensuremath{\,M_{\rm E}}}
\providecommand{\re}{\ensuremath{\,R_{\rm E}}}
\providecommand{\fave}{\langle F \rangle}
\providecommand{\fluxcgs}{10$^9$ erg s$^{-1}$ cm$^{-2}$}
\begin{table*}
\scriptsize
\centering
\caption{Median values and 68\% confidence intervals for the planetary parameters in the global model of HD~332231}
\begin{tabular}{llcccc}
  \hline
  \hline
Parameter & Description (Units) & Values & & & \\
\hline
~~~~$P$\dotfill &Period (days)\dotfill &$18.71204\pm0.00043$\\
~~~~$R_p$\dotfill &Radius (\rj)\dotfill &$0.867^{+0.027}_{-0.025}$\\
~~~~$M_p$\dotfill &Mass (\mj)\dotfill &$0.244\pm0.021$\\
~~~~$T_C$\dotfill &Time of conjunction (\bjdtdb)\dotfill &$2458692.25579^{+0.00077}_{-0.00078}$\\
~~~~$T_0^\dagger$\dotfill &Optimal conjunction time (\bjdtdb)\dotfill &$2458729.67987\pm0.00038$\\
~~~~$a$\dotfill &Semi-major axis (au)\dotfill &$0.1436^{+0.0032}_{-0.0033}$\\
~~~~$i$\dotfill &Inclination (degrees)\dotfill &$89.68^{+0.22}_{-0.28}$\\
~~~~$e$\dotfill &Eccentricity \dotfill &$0.032^{+0.030}_{-0.022}$\\
~~~~$\omega_*$\dotfill &Argument of periastron (degrees)\dotfill &$47^{+69}_{-74}$\\
~~~~$e\cos{\omega_*}$\dotfill & \dotfill &$0.011^{+0.028}_{-0.018}$\\
~~~~$e\sin{\omega_*}$\dotfill & \dotfill &$0.010^{+0.034}_{-0.019}$\\
~~~~$T_{eq}^\ddagger$\dotfill &Equilibrium temperature (K)\dotfill &$876\pm17$\\
~~~~$\tau_{\rm circ}$\dotfill &Tidal circularization timescale (Gyr)\dotfill &$630^{+110}_{-100}$\\
~~~~$K$\dotfill &RV semi-amplitude (m s$^{-1}$)\dotfill &$17.3\pm1.2$\\
~~~~$R_P/R_*$\dotfill &Radius of planet in stellar radii \dotfill &$0.06976^{+0.00041}_{-0.00039}$\\
~~~~$a/R_*$\dotfill &Semi-major axis in stellar radii \dotfill &$24.21^{+0.62}_{-0.78}$\\
~~~~$\delta$\dotfill &Transit depth (fraction)\dotfill &$0.004867^{+0.000058}_{-0.000055}$\\
~~~~$\tau$\dotfill &Ingress/egress transit duration (days)\dotfill &$0.01700^{+0.00086}_{-0.00029}$\\
~~~~$T_{14}$\dotfill &Total transit duration (days)\dotfill &$0.2565^{+0.0011}_{-0.0010}$\\
~~~~$b$\dotfill &Transit impact parameter \dotfill &$0.133^{+0.12}_{-0.092}$\\
~~~~$\rho_p$\dotfill &Density (g cm$^{-3}$)\dotfill &$0.464^{+0.054}_{-0.052}$\\
~~~~$\log{g_p}$\dotfill &Surface gravity \dotfill &$2.905^{+0.040}_{-0.044}$\\
~~~~$\fave$\dotfill &Incident flux (\fluxcgs)\dotfill &$0.1335^{+0.011}_{-0.0098}$\\
~~~~$T_P$\dotfill &Time of periastron (\bjdtdb)\dotfill &$2458690.2^{+3.3}_{-3.9}$\\
~~~~$T_S$\dotfill &Time of eclipse (\bjdtdb)\dotfill &$2458701.74^{+0.34}_{-0.21}$\\
~~~~$T_A$\dotfill &Time of ascending node (\bjdtdb)\dotfill &$2458687.72^{+0.27}_{-0.16}$\\
~~~~$T_D$\dotfill &Time of descending node (\bjdtdb)\dotfill &$2458696.93^{+0.20}_{-0.21}$\\
~~~~$\dot{\gamma}$\dotfill &RV slope (m s$^{-1}$ day$^{-1}$)\dotfill &$0.023\pm0.020$\\
\smallskip\\\multicolumn{2}{l}{Wavelength Parameters}&TESS\smallskip\\
~~~~$u_{1}$\dotfill &Linear limb-darkening coefficient \dotfill &$0.253\pm0.027$\\
~~~~$u_{2}$\dotfill &Quadratic limb-darkening coefficient \dotfill &$0.289\pm0.034$\\
\smallskip\\\multicolumn{2}{l}{Telescope Parameters}&APF&KECK&SONG\smallskip\\
~~~~$\gamma_{\rm rel}$\dotfill &Relative RV offset (m s$^{-1}$)\dotfill &$1.3\pm1.2$&$-1.9\pm1.2$&$1.9\pm6.5$\\
~~~~$\sigma_J$\dotfill &RV jitter (m s$^{-1}$)\dotfill &$8.87^{+1.0}_{-0.92}$&$3.64^{+1.4}_{-0.99}$&$19.8^{+7.1}_{-5.2}$\\
\hline
\end{tabular}
 \begin{flushleft} 
  \footnotesize{ 
    \textbf{\textsc{Notes.}}
See Table~3 in \citet{Eastman2019} for the definition and explanation of the derived and fitted parameters in \texttt{EXOFASTv2}.\\
$^\dagger$Minimum covariance with period.\\
$^\ddagger$Equilibrium temperature is calculated assuming zero albedo and perfect heat redistribution.}
\end{flushleft}
\label{tab:exofastv2}
\end{table*}


\section{Discussion}\label{sec:disc}

\subsection{A Planet on the Border between Neptunian and Jovian Classes}

\begin{figure}
    \centering
    \includegraphics[width=\columnwidth]{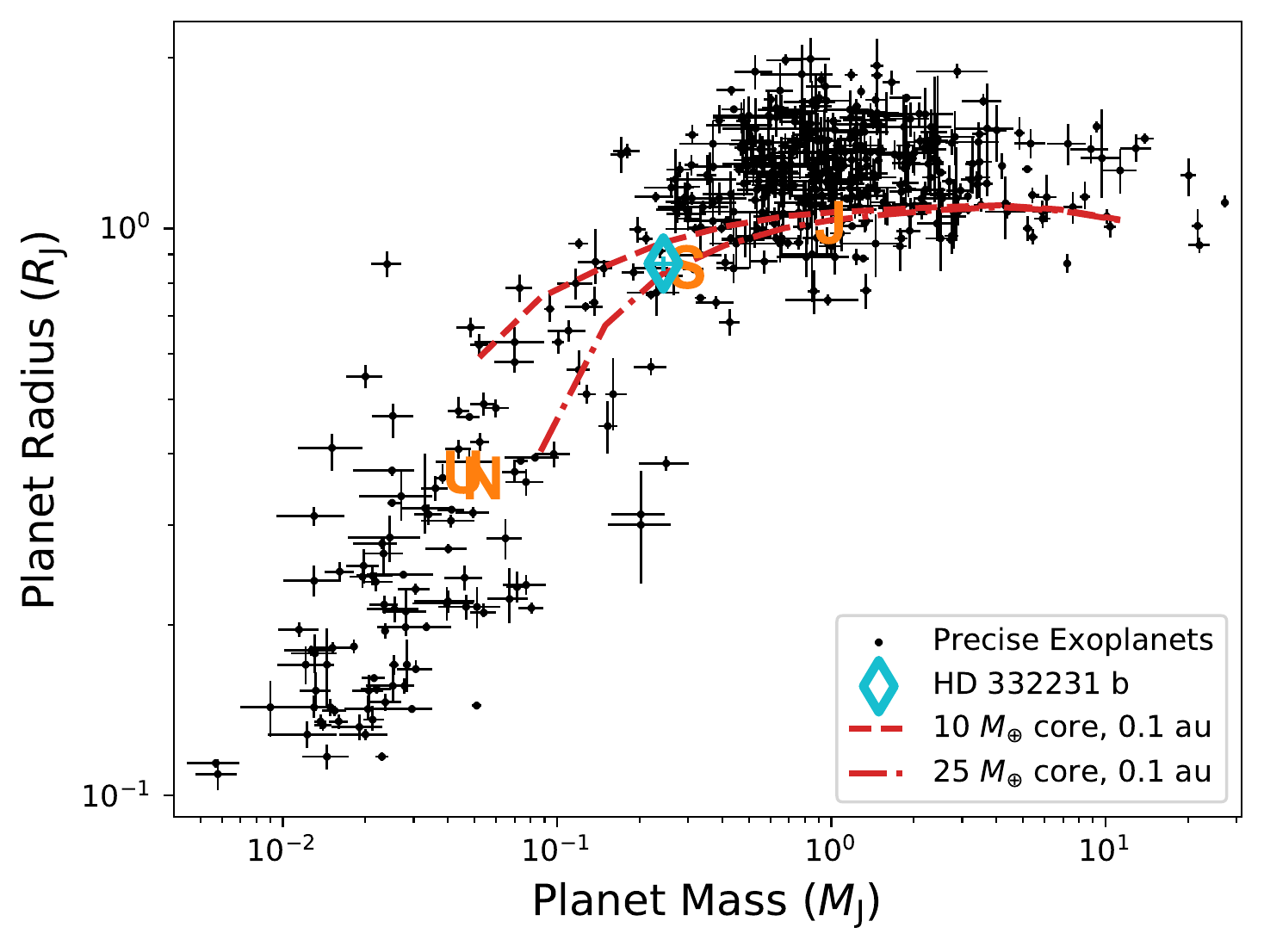}
    \caption{All known exoplanets with mass and radius measured to better than 30\% precision (according to the NASA Exoplanet Archive as of 2019 December 17). The solar system's giant planets are indicated by the letters ``J,'' ``S,'' ``U,'' and ``N.'' The dashed and dashed--dotted lines show the masses and radii of hydrogen--helium-dominated planets on 0.1~au orbits with core masses of 10 and 25~$M_{\earth}$, respectively \citep{Fortney2007}. The error bars for \hdb\ are shown but are smaller than the marker.}
    \label{fig:mp_v_rp}
\end{figure}

\hdb\ occupies an interesting area of parameter space relative to the solar system planets. By radius, \hdb\ is 13\% smaller than Jupiter and 4\% larger than Saturn. Yet its mass is 19\% below that of Saturn. Following the planetary class divisions of \citet{Chen2017}, the mass of \hdb\ places it among ``Neptunian'' planets as opposed to ``Jovian'' planets. Indeed, on a mass-radius diagram of well characterized planets (Figure \ref{fig:mp_v_rp}), \hdb\ resides on the edge of the cluster of giant planets whose radii are seemingly not correlated with planet mass. This makes \hdb\ an interesting case for further exploration into the division between Neptunian and Jovian planets.

To explore the possible interior properties of \hdb, we overlay contours in Figure \ref{fig:mp_v_rp} showing the core mass of giant hydrogen--helium-rich planets with 0.1~au orbits according to planetary evolution models \citep{Fortney2007}. These models suggest a core mass for \hdb\ between 10 and 25~$M_{\earth}$, which would account for 13--32\% of its total mass. This prediction is broadly consistent with current estimates for the core masses of Jupiter \citep[7--25 $M_{\earth}$;][]{Wahl2017} and Saturn \citep[5--20 $M_{\earth}$;][]{Helled2013}. Although 10--25~$M_{\earth}$ is a wide range, the planetary evolution models of \citet{Fortney2007} clearly predict that the core of \hdb\ is unlikely to be 0~$M_{\earth}$ (i.e., no core).

If the core mass of \hdb\ is similar to those of Jupiter and Saturn, then could \hdb\ be a Jovian planet with a relatively low-mass envelope? This possibility could be explored through a measurement of the metal enrichment in the atmosphere of \hdb. The atmospheres of Jupiter and Saturn are enriched in heavy elements relative to solar composition gas by factors of $\sim$4 and $\sim$10, respectively \citep{Wong2004,Fletcher2009}. This enrichment relates to the accretion of planetesimals during formation, which in turn depends on when and where formation occurred, the properties of the protoplanetary disk, and the presence of companions. If the atmosphere of \hdb\ is not enriched in metals, this could explain the planet's relatively low mass compared to the solar system planets and could lend insight into its formation history. Alternatively, if \hdb\ is indeed more like a Neptunian planet, as suggested by the class divisions of \citet{Chen2017}, then we might expect greater heavy element enrichment as is seen for Uranus and Neptune \citep{Sromovsky2011,Karkoschka2011}.  

\begin{figure}
    \centering
    \includegraphics[width=\columnwidth]{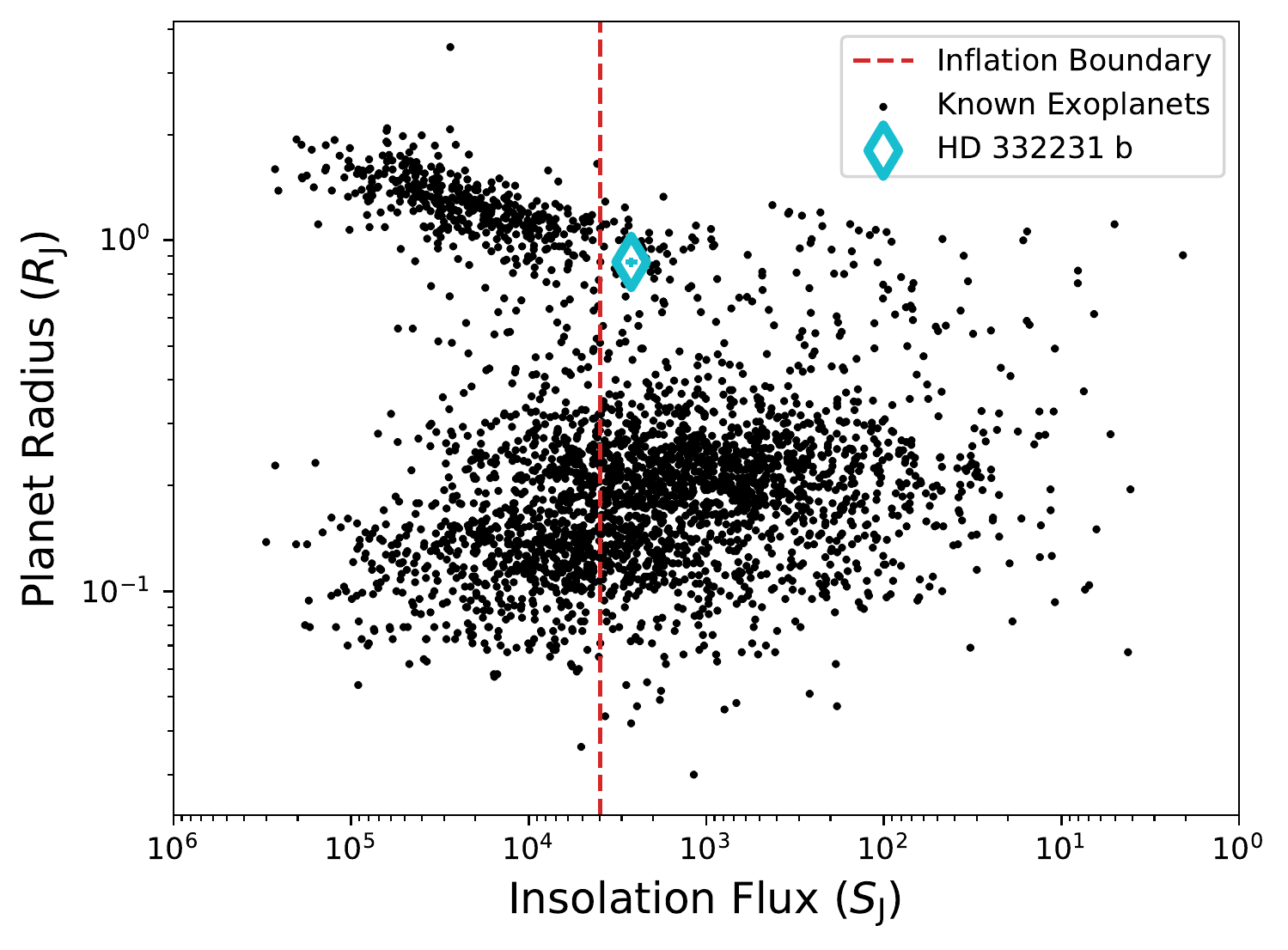}
    \caption{Planetary radius ($R_p$) versus insolation flux ($S$) for confirmed exoplanets with the planetary and stellar parameters needed to calculate $S$ (according to the NASA Exoplanet Archive as of 2019 December 17). The error bars for \hdb\ are shown but are smaller than the marker. The amount of flux \hdb\ receives is below the empirical radius inflation boundary of \citet{Demory2011b}, suggesting that its radius is not inflated. \hdb\ joins the minority of giant exoplanets with measured radii that are likely not inflated.}
    \label{fig:rp_v_s}
\end{figure}

The bulk density of \hdb\ is $0.464^{+0.054}_{-0.052}$~g~cm$^{-3}$, 32\% below that of Saturn. We consider whether this relatively low density is a result of planetary radius inflation. Highly irradiated exoplanets have been found to have larger radii than predicted by standard planetary evolution models, which has been attributed to several possible mechanisms \citep[see, for instance,][]{Fortney2010b,Baraffe2014}. However, empirical evidence suggests that the mechanisms causing radius inflation are not active for giant planets receiving an insolation flux ($S$) below $\sim$ 2\e{8}~erg~s$^{-1}$~cm$^{-2}$ \citep{Demory2011b}. Based on the global fit (described in Section \ref{sec:analysis}), we find the insolation flux of \hdb\ to be $S =$~1.34\e{8}~erg~s$^{-1}$~cm$^{-2}$, slightly below the inflation boundary. In Figure \ref{fig:rp_v_s}, we place this value in the context of other known exoplanets with measured radii. Indeed, \hdb\ occupies a region of uninflated giants planets away from the sharp upward ramp in radius. Of the known giant exoplanets with measured radii, \hdb\ is one of the minority that receives a low enough insolation flux to be uninflated by mechanisms relating to incident radiation. On its own, this makes \hdb\ a valuable discovery. 

This argument has several caveats. First, the radius inflation boundary is largely based on radius measurements of Jovian exoplanets \citep{Demory2011b}. If this boundary contains some yet unknown dependence on planetary mass, our conclusion that \hdb\ is not inflated may need to be revisited. However, at present, too few Saturn-mass exoplanets are known to investigate possible variations in the inflation boundary. Second, it is possible that high-altitude aerosols or hazes could be affecting the radius of \hdb\ \citep[e.g.,][]{Lammer2016,Gao2020}. This caveat could possibly be tested through atmospheric characterization.  

\subsection{Atmospheric Characterization Prospects}

Metal enrichment in exoplanetary atmospheres can be probed via transmission spectroscopy, for which \hdb\ is a strong candidate. Using the properties of \hdb\ determined in this work, we find that the transmission spectroscopy metric \citep[TSM;][]{Kempton2018} for \hdb\ is TSM $=216$. This calculation, which includes a scale factor of 1.15 for sub-Jovians, suggests that a 10~hr observation with the Near Infrared Imager and Slitless Spectrograph (NIRISS) on the James Webb Space Telescope would achieve an S/N of approximately 216 \citep{Louie2018}. This value readily surpasses the TSM$=$96 recommended threshold \citep{Kempton2018} and demonstrates that \hdb\ is a promising target for atmospheric characterization. 

Additionally, using the TSM value of \hdb\ as a proxy for relative observability, we note that, among the planets with similar properties, \hdb\ stands out as the most promising target. We compute the TSM value for all known transiting planets and TOIs (using the empirical mass-radius relation of \citet{Chen2017} as implemented in \citet{Louie2018} if masses are unknown) around late F- and early G-type stars (5500~K $< T_{\rm eff} <$ 6500~K) with radii 6.3~$R_{\earth} < R_{\rm p} <$ 11.2~$R_{\earth}$ and insolation flux values 10$^2$--10$^3$ times that of Earth. Of planets satisfying these criteria, \hdb\ has the highest TSM value, with the next-best being  TOI~1135.01, which is one magnitude fainter in the $V$-band.

With an equilibrium temperature of 876~K, \hdb\ could add dynamic range to studies of atmospheric properties as a function of temperature \citep[e.g.,][]{Stevenson2016} and would serve as a stepping stone between transmission spectroscopy of hot \citep[e.g.,][]{Sing2016} and cold \citep{Dalba2015} giant planets.   

Table \ref{tab:exofastv2} includes the transit ephemeris of \hdb\ measured from the \tess\ photometry. The precision on predictions of future transit times will degrade as time passes and future transits are not observed. From the ephemeris presented here, we estimate that the uncertainty on mid-transit time will rise to 3.0~m and 8.8~m by the years 2021 and 2030, respectively. This uncertainty is unlikely to affect future efforts to characterize the atmosphere of \hdb\ during transit. However, since only two transits of \hdb\ have been observed, it is not known if the ephemeris exhibits transit timing variations (TTVs). Based on the typical magnitude of TTVs in other transiting systems \citep[][e.g.,]{Holczer2016}, it is unlikely that TTVs for \hdb\ (if any) would preclude future in-transit atmospheric characterization \citep[e.g.,][]{Dalba2016,Dalba2019c}. Still, the discovery of TTVs could potentially identify the existence of a companion (if any) in the \hd\ system. For this reason, it may be useful to observe an additional transit of \hdb\ in the near future.

\subsection{Considerations of Planetary Companions and Formation History}\label{sec:disc_comp}

Another possible explanation for the mass of \hdb\ relative to its radius is the existence of a companion that reduced the amount of material in its feeding zone or truncated its growth during formation. The presence of an additional Keplerian signal (caused by a companion) is not evident in the RV time series or residuals presented in Section \ref{sec:data_rv} (Figure \ref{fig:rvs}). However, it is possible that the RV time series data show a slight linear trend that could be due to an outer companion with an orbital period much longer than the baseline of observations. We used the model comparison feature of \texttt{RadVel} to explore this possibility. When allowed to fit for a first-order acceleration term ($\dot \gamma$), \texttt{RadVel} converged to a solution with $\dot \gamma = 0.028\pm0.022$~m~s$^{-1}$~days$^{-1}$. We investigated the significance of this finding by comparing the Bayesian information criteria \citep[BIC;][]{Schwarz1978} and the Akaike information criteria \citep[AIC;][]{Akaike1974}, corrected for a small sample size \citep[AIC$_{\rm c}$;][]{Sugiura1978}, for models with or without $\dot \gamma$. The models were nearly indistinguishable from each other, having $\Delta$BIC = 2.13 and $\Delta$AIC$_{\rm c}$ = 0.22 in favor of the model without the acceleration parameter. 

While the tentative trend in the RVs described by the $\dot \gamma$ parameter could have other explanations (e.g., stellar variability), it is possible that a longer RV baseline for \hd\ could yield the discovery of a companion. RV and transit surveys have shown that the presence of outer giant planets is likely correlated to the existence of smaller inner planets \citep{Bryan2019,Herman2019,Zhu2018}. In these studies, ``smaller'' typically means the inner planets are intermediate in size between the Earth and Neptune. \hdb\ would be an interesting case if it has a companion, as it is the size of a Saturn but its low mass could qualify it as a Neptunian planet \citep{Chen2017}.

The determination of additional planets in this system (or lack thereof) would also provide valuable context to the formation history of \hdb. With an orbital period of $\sim$18~days, it is unlikely that this planet experienced any tidal circularization\footnote{The tidal circularization timescale for \hdb\ is found to be $\sim$630~Gyr (Table \ref{tab:exofastv2}).}, which all but rules out inward migration by mechanisms invoking high-eccentricity excitement. The possible existence of an outer companion may perhaps challenge this in the future. Alternatively, \hdb\ could have migrated via interactions with its protoplanetary disk or formed in situ \citep{Huang2016}. The latter case would be supported if any small, nearby companions were to be found also transiting in the \hd\ system. We do not find obvious evidence for transits of any such objects, although we have neither conducted an exhaustive search nor quantified our detection limits with an injection-recovery test. Therefore, \hdb\ may demonstrate that disk migration can place giant planets on relatively short orbits.  

The orbital period of \hdb\ places it squarely in the ``period valley,'' which is a proposed decrease in the occurrence rate of giant planets with orbital periods between roughly 10 and 100~days \citep[e.g.,][]{Udry2003}. Different observational methods have yielded different conclusions regarding the existence and extent of the valley in period space \citep[e.g.,][]{Udry2007,Howard2012}. However, both RV and transit surveys seem to agree that it is present between orbital periods of 10--20~days \citep{Santerne2016,Santerne2018}. The $\sim18.7$-day orbital period of \hdb\ may then require special explanation. One possible explanation is that a companion halted the inward migration of \hdb\ at some point either during or after its formation. This explanation has also been proposed to explain why Jupiter itself avoided a much shorter orbit during the solar system's formation \citep{Morbidelli2007b}. Continued RV monitoring of \hd\ can readily test this hypothesis by searching for an outer companion.    

\subsection{Prospects for Measuring the Stellar Obliquity}

The brightness and projected rotation velocity of \hd\ combined with the transit depth of \hdb\ makes the system amenable to a measurement of the stellar obliquity via the Rossiter--McLaughlin (RM) effect \citep[e.g.,][]{Winn2005}. We estimate that the amplitude of the RM effect is roughly 20~m~s$^{-1}$ assuming that $v \; \sin{i}=6$~km~s$^{-1}$. The nearly central transit (i.e., impact parameter consistent with zero) may, however, challenge a spin-orbit measurement since it introduces a degeneracy into the model \citep{Gaudi2007}. Given the timing of the RV observations and the duration of transit, it is unlikely that any of the RV measurements presented here are affected by this effect. 

At the orbital separation of \hdb, tidal dissipation is unlikely to have lowered the obliquity from its initial value. This means that \hdb\ may have maintained its obliquity and may provide a window to the initial distribution of giant planet obliquities.

\subsection{Scattered Light and Data Masking}

Observations made by \tess\ in Sectors 14, 15, and 16 suffered from contamination from scattered light due to the Earth and Moon. The scattered light signal differs between each detector and even across a single detector. As described in the data release notes accompanying the \tess\ data\footnote{\url{https://archive.stsci.edu/tess/tess_drn.html}}, the cadences suffering from scattered light would adversely affect the systematic error removal in the data conditioning and also the planet search. As a result, these data points were masked in the PDC light curves (Section \ref{sec:data_tess}). 

Here, we briefly consider the consequence that this scattered light contamination may have on a transit search in the PDC light curves. We acquired every two-minute cadence PDC light curve from Sectors 14, 15, and 16 from MAST\footnote{Accessed 2019 November 21 (\url{https://archive.stsci.edu/tess/}).}. We scanned each light curve for quality flags that identified masked data due to scattered light. Of the combined cadences for all PDC light curves in Sector 14, 5.4\% are masked due to scattered light. This fraction is 5.2\% and 1.7\% for Sectors 15 and 16, respectively. 

\begin{figure}
    \centering
    \includegraphics[width=\columnwidth]{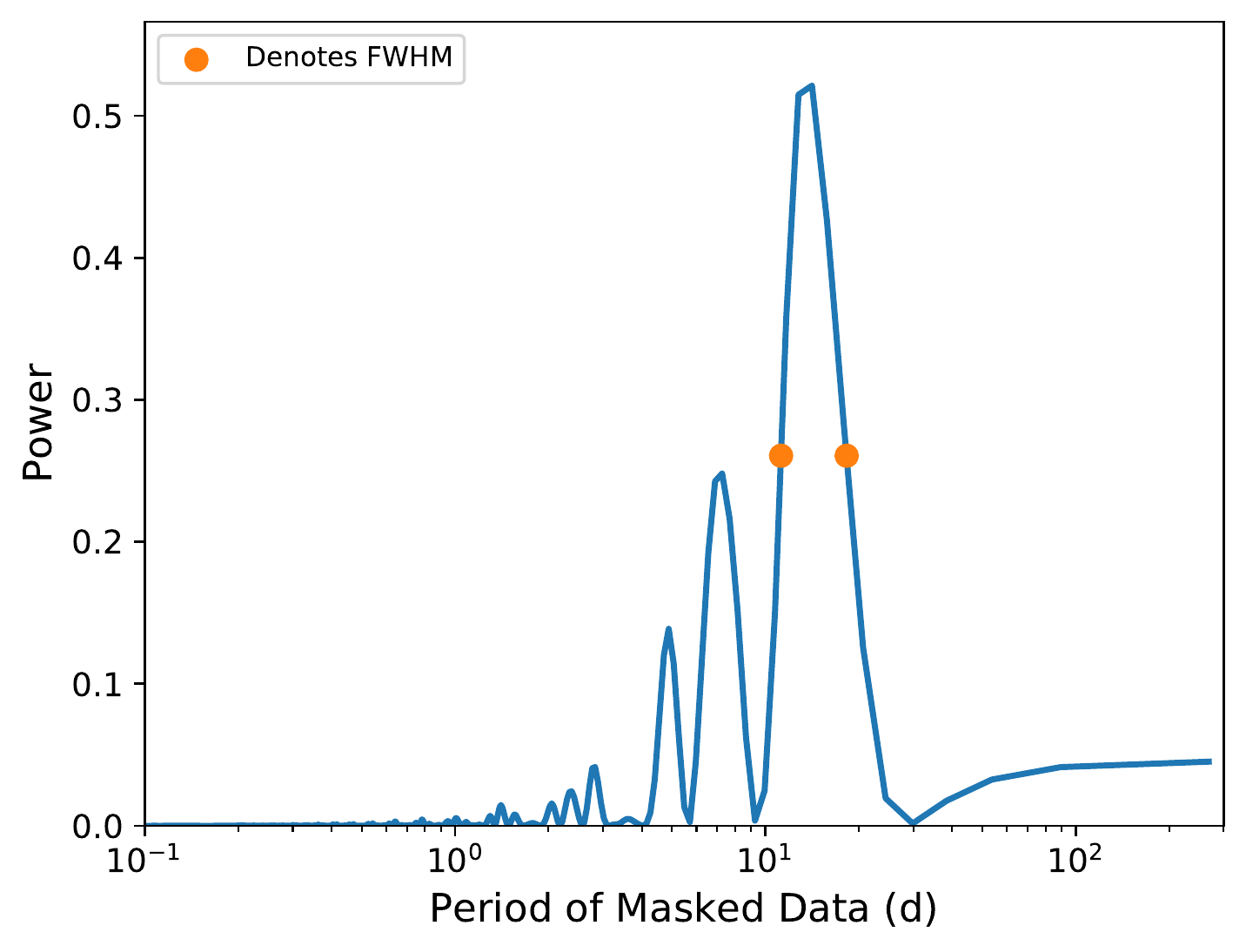}
    \caption{Periodogram of the spectral window of the masked data for \hd\ from Sector 14. Peaks due to the data cadence are not shown. This periodogram shows a peak near 14~days with a FWHM of $\sim$7~days, which is typical for stars observed in Sector 14 at two-minute cadence.}
    \label{fig:window}
\end{figure}

The timing of the data masking is not random. As described in the \tess\ data release notes, the masking can result in data gaps of several days. We calculated the periodogram of the spectral window of the masked data for each of the PDC light curves in Sectors 14, 15, and 16. Peaks due to the two-minute cadence of the data were ignored. The periodogram of the spectral window of the masked data for \hd\ in Sector 14 is shown in Figure \ref{fig:window} as an example. In the Sector 14 light curves, the peaks in the periodogram occur over the narrow range of 14.1313--14.1318~days, but the full width at half maximum (FWHM) power of each of these signals is 7.0--7.1~days. Unsurprisingly, this periodicity matches the orbital period of \tess\ around the Earth. The 7-day FWHM, however, demonstrates that all transits of exoplanets with periods in the range 10.5--17.5~days could be hidden by the masking in the PDC light curves if only observed in Sector 14. Results are similar for Sectors 15 and 16 where the masked data have periodicities of $\sim$13.7~days and $\sim$14.5~days and FWHM values of $\sim$7.4~days and $\sim$6.8~days, respectively. 

Inspecting the target pixel data (as in Section \ref{sec:data_tess}) for individual Sectors 14, 15, and 16 targets is straightforward and highly advisable. However, such inspection may be more challenging for comprehensive transit searches that benefit from the processed PDC light curves. Moving forward, it may be beneficial to develop additional means of addressing scattered light contamination in PDC light curves.


\section{Conclusion}\label{sec:conc}

The Transiting Exoplanet Survey Satellite (\tess) observed \hd\ (\toi), an eighth magnitude F8 dwarf star, in Sectors 14 and 15 of its primary mission (Figure \ref{fig:transit}). We identified a 6.2~hr, $\sim$4900 ppm single transit in the Sector 15 two-minute cadence PDC light curve of \hd. Subsequent spectroscopic and imaging observations from TRES (Section \ref{sec:data_recon}) and NESSI (Figure \ref{fig:speckle}) ruled out common non-planetary (i.e., false-positive) scenarios for this signal. We then began follow-up radial velocity observations with the APF telescope at Lick Observatory and the Keck I telescope at W. M. Keck Observatory (Figure \ref{fig:rvs}). The determination of the orbital period of the planet candidate via RV observations motivated the analysis of \tess\ target pixel data, which uncovered a Sector 14 transit that had previously been masked due to scattered light contamination (Figure \ref{fig:S14_scattered}). A joint analysis of the photometric and spectroscopic observations of \hd\ confirm the existence of a Saturn-size ($0.867^{+0.027}_{-0.025} \; R_{\rm J}$), sub-Saturn-mass ($0.244\pm0.021 \; M_{\rm J}$) exoplanet, \hdb.  

The orbit of \hdb\ has a period of 18.71~days and is essentially circular ($e=0.032^{+0.030}_{-0.022}$). Its low mass relative to its size places \hdb\ near the transition between ``Neptunian'' and ``Jovian'' regimes \citep{Chen2017}. We use planetary evolution models to infer possible interior properties of \hdb\ and place it in context with other exoplanets and the solar system giant planets (Figures \ref{fig:mp_v_rp} and \ref{fig:rp_v_s}). We predict that atmospheric characterization---especially a measurement of its atmospheric metal enrichment---would be useful in characterizing \hdb. The TSM \citep{Kempton2018} for \hdb\ is 216, which makes it a promising weakly irradiated, giant planet target.  

We do not find evidence for a companion in the \hd\ system. However, we advocate for continued spectroscopic monitoring of the system in search of additional Keplerian signals in the RV time series. The presence of a companion may help to explain the properties of \hdb\ including its existence in the proposed ``period valley'' \citep[e.g.,][]{Santerne2016}. 

Finally, we conduct a brief investigation of the scattered light in Sectors 14--16 that suggests that transits of exoplanets with orbital periods between 10.5 and 17.5~days could be hidden due to data masking in \tess\ PDC light curves.


\acknowledgements

The authors thank the anonymous referee for a helpful review of this work. P.D. gratefully acknowledges support from a National Science Foundation Astronomy \& Astrophysics Postdoctoral Fellowship under award AST-1903811. E.A.P. acknowledges the support of the Alfred P. Sloan Foundation. L.M.W. is supported by the Beatrice Watson Parrent Fellowship and NASA ADAP Grant 80NSSC19K0597. A.C. acknowledges support from the National Science Foundation through the Graduate Research Fellowship Program, grant No. DGE 1842402. D.H. acknowledges support by the National Aeronautics and Space Administration (80NSSC18K1585, 80NSSC19K0379) awarded through the TESS Guest Investigator Program. I.J.M.C. acknowledges support from the NSF through grant AST-1824644. A.B. is supported by the NSF Graduate Research Fellowship, grant No. DGE 1745301. R.A.R. is supported by the NSF Graduate Research Fellowship, grant No. DGE 1745301. C.D.D. acknowledges the support of the Hellman Family Faculty Fund, the Alfred P. Sloan Foundation, the David \& Lucile Packard Foundation, and the National Aeronautics and Space Administration via the TESS Guest Investigator Program (80NSSC18K1583). M.R.K is supported by the NSF Graduate Research Fellowship, grant No. DGE 1339067. J.M.A.M. thanks the LSSTC Data Science Fellowship Program, which is funded by LSSTC, NSF Cybertraining Grant No. 1829740, the Brinson Foundation, and the Moore Foundation; his participation in the program has benefited this work. J.M.A.M. also gratefully acknowledges support from the National Science Foundation Graduate Research Fellowship under Grant No. DGE-1842400. R.C. is supported by a grant from the National Aeronautics and Space Administration in support of the TESS science mission. The contribution by M.L. was carried out within the framework of the National Centre for Competence in Research ``PlanetS'' supported by the Swiss National Science Foundation (SNSF). M.L. also acknowledges support from the Austrian Research Promotion Agency (FFG) under project 859724 ``GRAPPA.'' T.D. acknowledges support from MIT's Kavli Institute as a Kavli postdoctoral fellow. D.D. acknowledges support from NASA through Caltech/JPL grant RSA-1006130 and through the TESS Guest Investigator Program Grant 80NSSC19K1727. D.J.S. is supported as an Eberly Research Fellow by the Eberly College of Science at the Pennsylvania State University. The Center for Exoplanets and Habitable Worlds is supported by the Pennsylvania State University, the Eberly College of Science, and the Pennsylvania Space Grant Consortium. H.P.O. acknowledges funding from NCCR-PlanetS for the CHESS postdoctoral fellowship. 

The authors wish to thank Bradford Holden for assistance in scheduling some of the observations presented in this paper. T.J., M.H.K., and D.L. gratefully acknowledge Allan R. Schmitt and Troy Winarski for making their light-curve software \texttt{LcTools} and \texttt{AKO-TPF} freely available.  

Resources supporting this work were provided by the NASA High-End Computing (HEC) Program through the NASA Advanced Supercomputing (NAS) Division at Ames Research Center for the production of the SPOC data products.

This research has made use of the NASA Exoplanet Archive, which is operated by the California Institute of Technology, under contract with the National Aeronautics and Space Administration under the Exoplanet Exploration Program.

This research has made use of the Exoplanet Follow-up Observation Program website, which is operated by the California Institute of Technology, under contract with the National Aeronautics and Space Administration under the Exoplanet Exploration Program.

Funding for the Stellar Astrophysics Centre is provided by The Danish National Research Foundation (Grant agreement no.: DNRF106).

This paper includes data collected with the \tess\ mission, obtained from the MAST data archive at the Space Telescope Science Institute (STScI). Funding for the \tess\ mission is provided by the NASA Explorer Program. STScI is operated by the Association of Universities for Research in Astronomy, Inc., under NASA contract NAS 5–26555.

Observations in this paper made use of the NN-EXPLORE Exoplanet and Stellar Speckle Imager (NESSI). NESSI was funded by the NASA Exoplanet Exploration Program and the NASA Ames Research Center. NESSI was built at the Ames Research Center by Steve B. Howell, Nic Scott, Elliott P. Horch, and Emmett Quigley.

This paper includes observations made with the Hertzsprung SONG telescope operated on the Spanish Observatorio del Teide on the island of Tenerife by the Aarhus and Copenhagen Universities and by the Instituto de Astrof\'isica de Canarias. 

Research at the Lick Observatory is partially supported by a generous gift from Google. Some of the data presented herein were obtained at the W. M. Keck Observatory, which is operated as a scientific partnership among the California Institute of Technology, the University of California, and NASA. The Observatory was made possible by the generous financial support of the W.M. Keck Foundation.

Finally, the authors wish to recognize and acknowledge the very significant cultural role and reverence that the summit of Maunakea has always had within the indigenous Hawaiian community. We are most fortunate to have the opportunity to conduct observations from this mountain.

\vspace{5mm}
\facilities{Automated Planet Finder (Levy), FLWO:1.5m (TRES), Keck:I (HIRES), SONG, TESS, WIYN (NESSI).}
\software{\texttt{emcee} \citep{ForemanMackey2013}, \texttt{EXOFASTv2} \citep{Eastman2017}, \texttt{Lightkurve} \citep{Lightkurve2018}, \texttt{RadVel} \citep{Fulton2018}, \texttt{SpecMatch} \citep{Fulton2015b,Petigura2017}}

\end{document}